\documentclass[3p]{elsarticle}
\usepackage{lineno,hyperref}
\usepackage{amsmath}
\usepackage{caption}
\usepackage{subcaption}
\usepackage{indentfirst}
\usepackage{booktabs}
\usepackage{color}
\usepackage{transparent}
\usepackage{setspace}
\journal{Journal of Chemical Engineering Journal}

\begin{document}

\begin{frontmatter}

\title{Direct numerical simulation of  water-ethanol flows in a T-mixer}

\author[LSTM]{Tobias Schikarski}\ead{tobias.schikarski@fau.de}
\author[LFG]{Wolfgang Peukert}
\author[LSTM,ZARM]{Marc Avila}
\address[LSTM]{Institute of Fluid Mechanics, Department of Chemical and Biological Engineering, Friedrich-Alexander-Universit\"at Erlangen-N\"urnberg, 91058 Erlangen, Germany}
\address[LFG]{Institute of Particle Technology, Department of Chemical and Biological Engineering, Friedrich-Alexander-Universit\"at Erlangen-N\"urnberg, 91058 Erlangen, Germany}
\address[ZARM]{Center of Applied Space Technology and Microgravity, Universit\"at Bremen, Am Fallturm, 28359 Bremen, Germany}

%\author[mymainaddress,mysecondaryaddress]{Tobias Schikarski}

\begin{abstract}

The efficient mixing of fluids is key in many applications, such as chemical reactions and nanoparticle precipitation. Detailed experimental measurements of the mixing dynamics are however difficult to obtain, and so predictive numerical tools are helpful in designing and optimizing many processes.  If two different fluids are considered, the viscosity and density of the mixture depend often nonlinearly on the composition, which makes the modeling of the mixing process particularly challenging. Hence water-water mixtures in simple geometries such as T-mixers have been intensively investigated, but little is known about the dynamics of more complex mixtures, especially in the turbulent regime. We here present a numerical method allowing the accurate simulation of two-fluid mixtures. Using a high-performance implementation of this method we perform direct numerical simulations resolving the spatial and temporal dynamics of water-ethanol flows for Reynolds numbers from 100 to 2000. The flows states encountered during turbulence transition and their mixing properties are discussed in detail and compared to water-water mixtures. 

\end{abstract}


\end{frontmatter}

%\linenumbers
%%%%%%%%%%%%%%
% SECTION
%%%%%%%%%%%%%%
\section{Introduction}

The bioavailability of pharmaceutical drugs is the degree to which a drug becomes available to the target tissue after administration. It is associated with the aqueous dissolution rate of the drug and determined by the size of its constituent particles. Nearly 70\% of the drugs developed in recent years show poor water solubility and hence manufacturing processes to produce monodisperse particles of the smallest possible size have been the focus of much research. These processes can be either top-down or bottom-up and usually deal with the production of nanocrystals containing mainly active pharmaceutical ingredient. Liquid antisolvent (LAS) precipitation is a promising, yet poorly understood, bottom-up process. \citet{Thorat20121} recently provided a review of experimental and numerical approaches  to LAS precipitation, highlighting its paramount importance for the pharmaceutical industry. Its advantages are simple up-scaling, economical production and the generation of crystals in the nanometer range. In LAS precipitation the drug is initially dissolved in an organic solvent such as ethanol, which is then mixed with an anti-solvent as water. In the mixing zone a thermodynamic driving force (supersaturation) arises and the drug precipitates into nanoparticles. Their size is determined by the supersaturation, which in turn strongly depends on the mixing efficiency. Consequently, a detailed understanding of the fluid dynamics of such mixing phenomena and its connection to particle formation is required to control and optimize the outcome of LAS.

\begin{figure}[t!]
  \begin{center}
    \begin{tabular}{cc}  
      \includegraphics[width=0.5\linewidth]{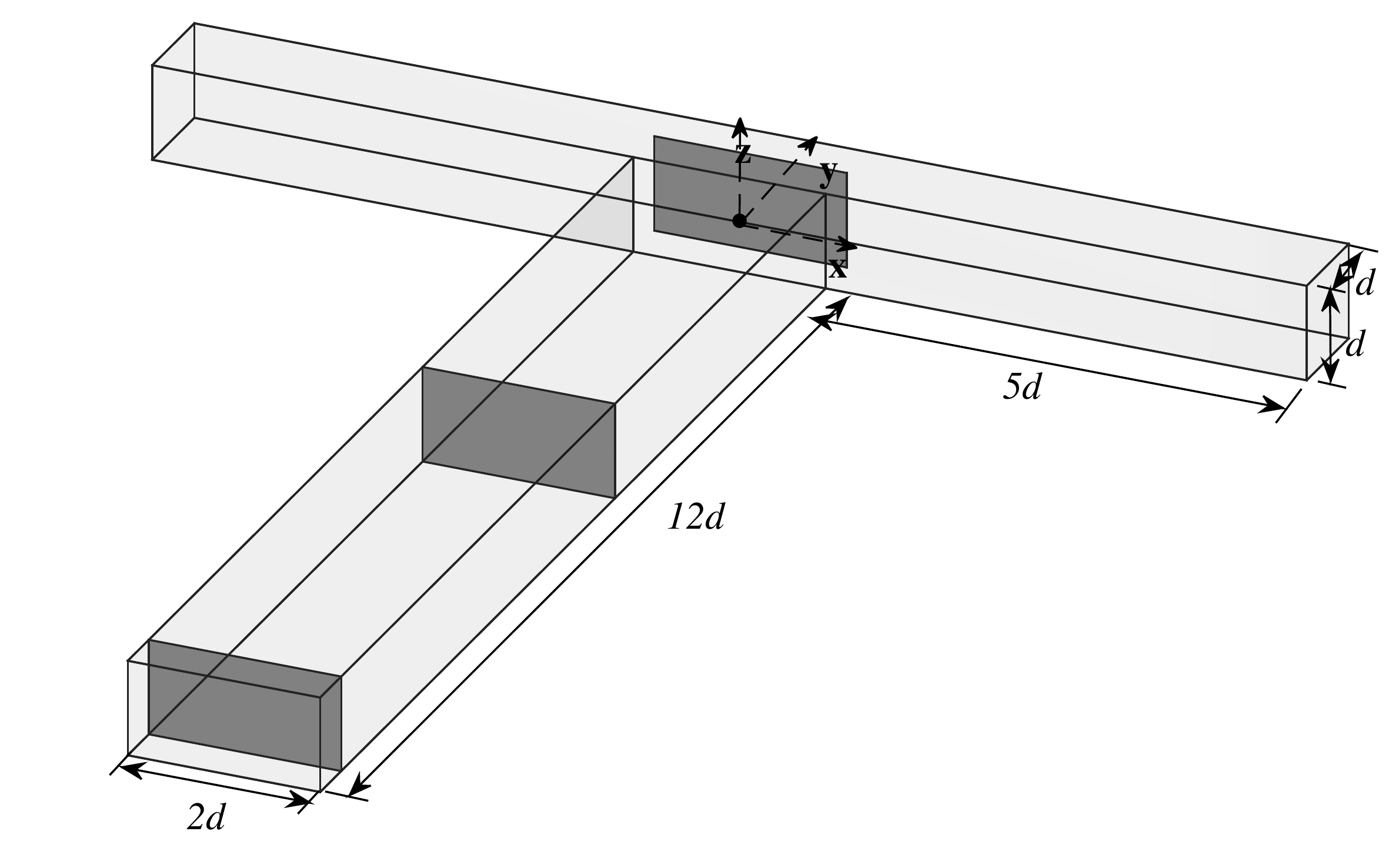} &
      \includegraphics[width=0.43\linewidth]{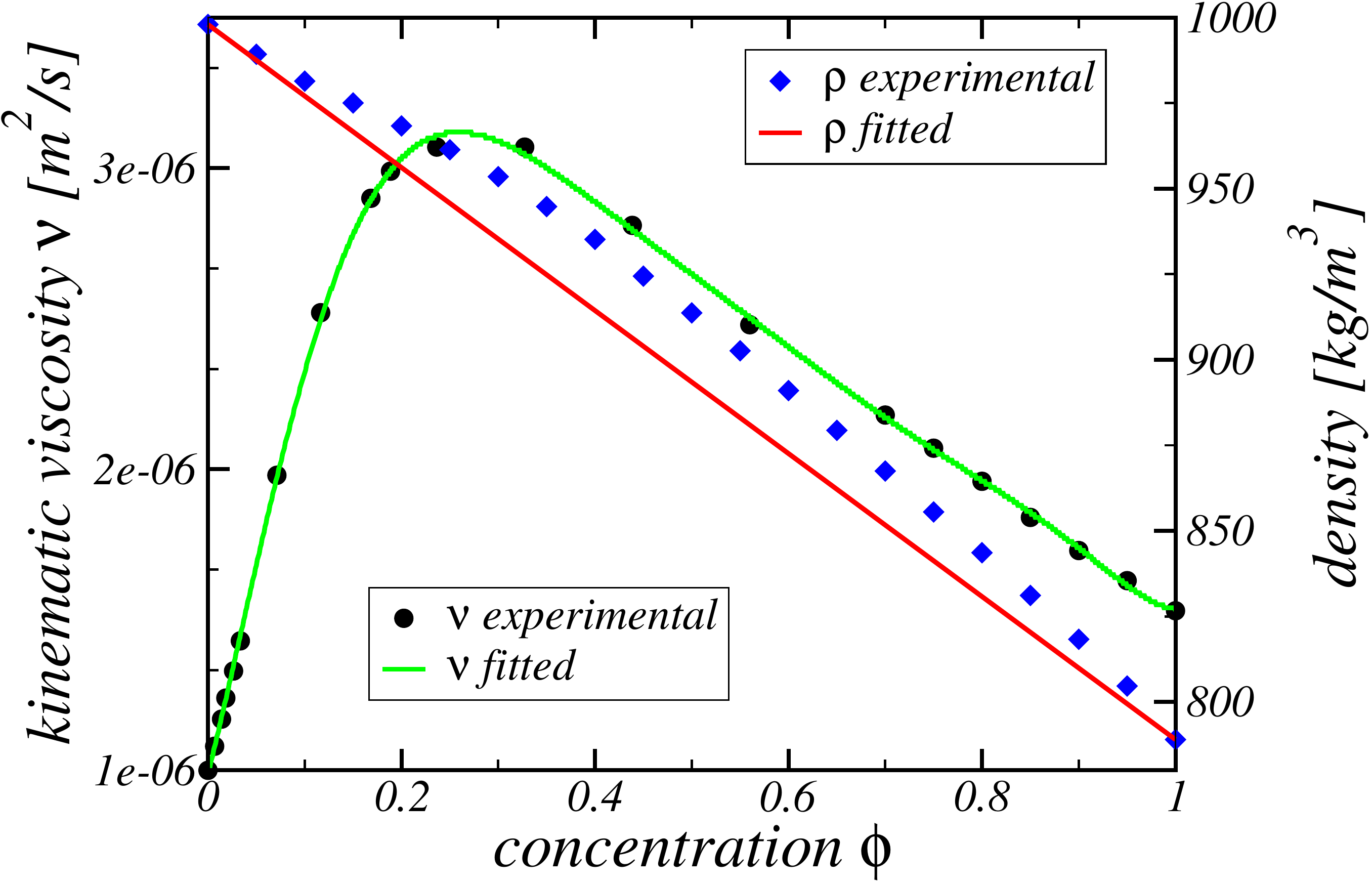}\\
      (a) & (b)
    \end{tabular}     
  \end{center}
\caption{(a) T-shaped mixer consisting of two quadratic inlet channels and a rectangular mixing outlet channel. The width and height of the inlet channel is $d=1$mm and the origin of the coordinate system is located at the middle of the mixing and inlet channel. The dark-shaded cross section at $y/d=(0,-5.5,-12)$ is used later to characterise the properties of the mixing process. (b) Kinematic viscosity $\nu$ (y-axis) and density $\rho$ (alt y-axis) as a function of the concentration $\phi$ for a water-ethanol mixtures. Here $\phi=0$ and $\phi=1$ correspond to the pure water and ethanol phases, respectively. The solid lines indicate functions fitted to the experimental data (dots and squares), see  \cite{dizechi}. }
\label{fig:TJunctionPFit}
\end{figure}

Despite the growing attention that LAS precipitation has attracted in the last decade, few works have   focused on the accurate modelling of the mixing of the solvent (alcohol) and the anti-solvent (water), see \cite{Orsi2013174,Wang2012252}. Water-water (WW) mixtures in laminar and slightly transient regimes in T-shaped micromixers have been intensively investigated \cite{Hoffmann20062968, Bothe20062950, Bothe20116424, Fani20136, Fani20147, Orsi2013174}. Fig.~\ref{fig:TJunctionPFit}(a) shows a schematic of a simple T-micromixer as usually used to investigate the physics of mixing processes. The most important parameter of the system is the Reynolds numbers $Re=\bar{u} d /\nu$, where $\bar{u}$ is the mean inlet velocity, $d$ the diameter of the inlet and $\nu$ the kinematic viscosity of the fluid. The Schmidt number $Sc=\nu/D$, where $D$ is the diffusion coefficient, quantifies the rate of momentum to mass transfer is typically large in fluids. The flow patterns and corresponding mixing behaviour have been thoroughly characterised as a function of $Re$, and an evident improvement in mixing efficiency has been observed for increasing the Reynolds number \cite{Hoffmann20062968, Bothe20062950, Bothe20116424, Fani20136, Fani20147, Orsi2013174}. In addition, direct numerical simulations (DNS) of turbulent mixing, complemented with the simulation of appropriate population balance equations, have succeeded in predicting the experimental outcome of the precipitation of inorganic compounds such as barium sulphate \cite{Gradl2006908,Schwertfirm20071429}.

\begin{figure}
  \begin{center}
     \begin{tabular}{ccc}
    	(a) \small{$y/d=-0.25$} & (b) \small{$y/d=-5.5$}& (c) \small{$y/d=-12$}\\
    	      	\includegraphics[width=0.25\linewidth]{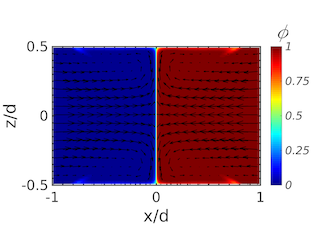} &
    	\includegraphics[width=0.25\linewidth]{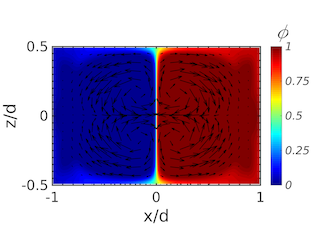} &
    	\includegraphics[width=0.25\linewidth]{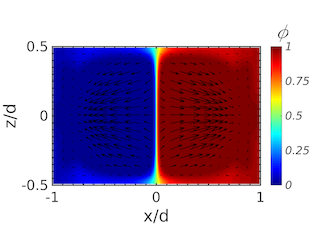}\\
      	\includegraphics[width=0.25\linewidth]{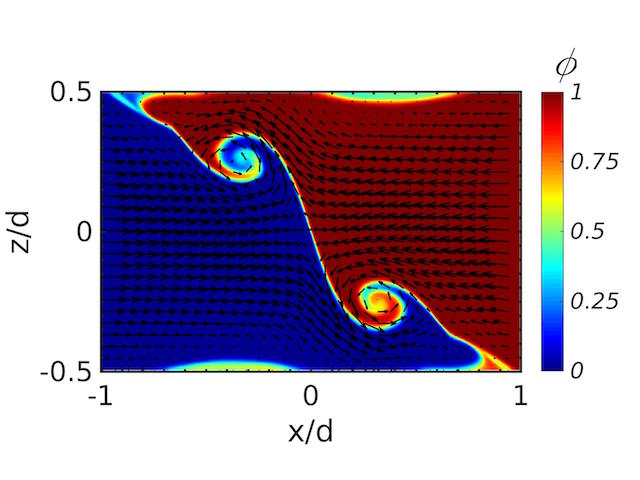} &
    	\includegraphics[width=0.25\linewidth]{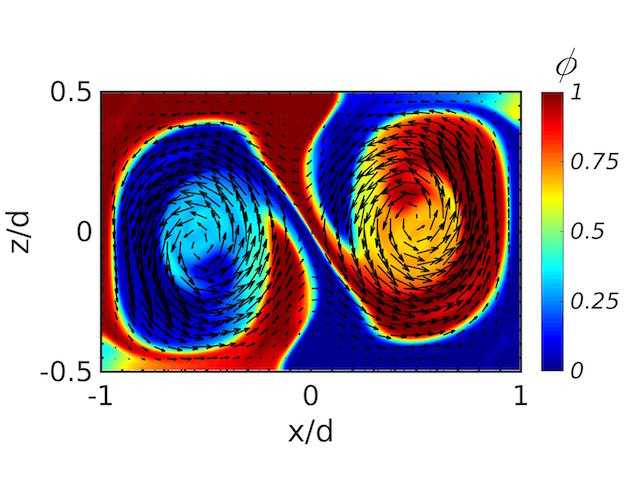} &
    	\includegraphics[width=0.25\linewidth]{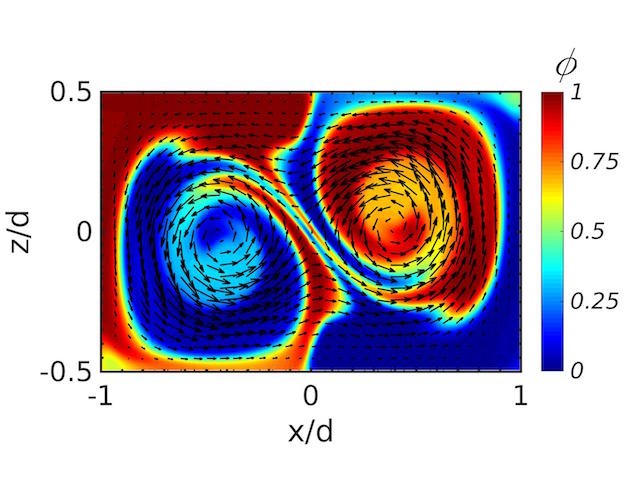}
     \end{tabular}
 \end{center}
  \caption{Colormap of the concentration $\phi$ at three cross-sections along the main channel for WW in the steady vortex regime ($Re=100$ and $Sc=600$, top row) and steady engulfment regime ($Re=160$ and $Sc=600$, bottom row). The arrows show the cross-sectional velocity field. At $Re=160$, the spatially-averaged mixing degree (see eq.~\ref{eqn:dom}) is $\delta =(0.041, 0.219, 0.243$), respectively, while at $Re=100$ the mixing degree is $\delta \approx 0$.}
\label{fig:WW160}
\end{figure}

At very low $Re$ the fluids streams flow side by side along the main channel. In the case that the concentration is a passive scalar, as in a water-water system, the velocity field and concentration field are top-down and left-right symmetric. Here mixing occurs only by diffusion at the straight contact plane and completely mixed state is reached after the distance $L\simeq \frac{d^2 \bar{u}}{D}$. However, $D$ is typically very small and thus in applications $L$ is prohibitively long. The simulations of \citet{ Bothe20062950, Bothe20116424} and experiments of \citet{Hoffmann20062968} showed that by increasing $Re$ vortex pairs arise at the junction, see the top panels in Fig.~\ref{fig:WW160}. However, in this so-called stratified regime the vortices are top-down and left-right symmetric. Hence a straight contact plane between the fluids persists and the mixing remains purely diffusive. By further increasing the Reynolds number they identified an additional steady regime, in which the left-right and top-down symmetries are broken and the resulting vortices intertwine the two inlet streams as shown in Fig.~\ref{fig:WW160}. In this engulfment regime the contact surface between the fluids becomes convoluted and henceforth the mixing efficiency is enhanced. Note that the engulfment flow remains symmetric with respect to a 180$^\circ$ rotation. \citet{Fani20136} analysed the instability mechanism and \citet{thomas2010experimental} experimentally reported that at even higher Reynolds numbers the flow becomes periodic and afterwards chaotic.  A stability analysis of these secondary transitions was carried out  by \citet{Fani20147}, who precisely determined the Reynolds numbers at which they occur. Both \citet{thomas2010experimental} as well as \citet{Fani20147} discussed qualitatively the enhancement of mixing because of transient flow. 

The first investigation of the mixing of water and ethanol (WE) was performed by \citet{Wang2012252}, who conducted $\mu$-LIF measurements in a T-mixer. By studying the LAS precipitation of curcumin they found a strong correlation of the mean particle size and the mixing efficiency, whereby a higher mixing quality resulted in a smaller mean particle size. Those findings motivated \citet{Orsi2013174} to precisely study the mixing of water with ethanol in the steady flow regime ($Re< 200$) with numerical simulation. They ascertained that flow regimes analogous to the water-water system are observed, however the transitions are hampered by the increase of viscosity as the two fluids mix. The strong dependence of the viscosity on the concentration of the mixture is shown in Fig.~\ref{fig:TJunctionPFit}(b).

\citet{Schwertfirm20071429} and \citet{Gradl2006908} demonstrated that DNS combined with appropriate models of the precipitation kinetics can quantitatively predict the precipitation of inorganic compounds. These elucidations outline the motivation of this work, namely to develop a code for DNS of miscible fluids with varying physical properties at operating Reynolds numbers. Such a code would be the first step toward a predictive tool for the outcome of LAS precipitation. Note however that real operating conditions for LAS precipitation are typically beyond $Re>2000$, which has not been reached with DNS for fluids with constant physical properties, let alone complex mixtures such as WE. The need for the DNS  arises from the naturally small molecular diffusion compared to the momentum diffusion, characterized by the Schmidt number $Sc\approx 600$, and the strong nonlinear change of the viscosity with concentration. Under these circumstances a direct application of LES and RANS additionally would bring along with the turbulent diffusion an unquantifiable parameter to this already challenging problem. Our work may serve as a benchmark for the development of industry accessible tools such as RANS and LES.

%%%%%%%%%%%%%%
% MODEL
%%%%%%%%%%%%%%
\section{Modeling assumptions and governing equations}
\label{sec:modeling}

We here consider miscible fluids with physical properties, such as density $\rho$ and dynamic viscosity $\mu$ that depend on the normalized concentration of the mixture, $\phi\in (0,1)$. Fitted functions based on experimental data \cite{dizechi} give smooth dependencies of the density $\rho(\phi)$ and the dynamic viscosity $\mu(\phi)$ as a function of the composition, see Fig.~\ref{fig:TJunctionPFit}(b). The transport of species (without chemical reactions) is modeled here with a convection-diffusion equation, which in conservative form reads
\begin{equation}
\frac{\partial}{\partial t} \int_V \rho(\phi)  \phi \, \mathrm{d}V = -\int_A \left( \rho(\phi) \boldsymbol{u} \phi \right) \cdot \boldsymbol{n} \, \mathrm{d}A + \int_A \left( \rho(\phi)  D \nabla \phi\right) \cdot \boldsymbol{n} \, \mathrm{d}A, \label{eqn:tes} 
\end{equation}
where $\boldsymbol{u} = (u_x,u_y,u_z)$ is the fluid velocity. The mixing process is considered to be isothermal, as justified in detail by \citet{Orsi2013174}. Furthermore, the volume excess naturally occurring while mixing ethanol and water is neglected. Although the density depends implicitly on time $\rho(\phi(t))$, we assume that the total mass is conserved. This is enforced by setting
\begin{equation} 
\int_A \left( \rho(\phi) \boldsymbol{u}) \right) \cdot \boldsymbol{n} \, \mathrm{d}A= 0 \label{eqn:mcd},
\end{equation}
which is known as the \textit{quasi-incompressibility} assumption. A detailed justification of the validity of this hypothesis can be found in \citet{joseph1990fluid,Joseph1996104}. The fluid motion is governed by the Navier--Stokes equations, which in conservative form read
\begin{equation} 
\frac{\partial}{\partial t} \int_V \rho(\phi)  \boldsymbol{u}  \, \mathrm{d}V  = - \int_A  \left( \rho(\phi)  \boldsymbol{u} \otimes \boldsymbol{u}\right) \cdot \boldsymbol{n}\, \mathrm{d}A - \int_V \nabla p  \, \mathrm{d}V + \int_A  \left( \mu(\phi) (\nabla \boldsymbol{u}+\nabla^T \boldsymbol{u}) \right) \cdot \boldsymbol{n} \, \mathrm{d}A. 
\label{eqn:tem} 
\end{equation}

The dynamics of the fluid motion is governed by the Reynolds number, defined as $Re=\frac{\bar{u}d}{\nu_w}$, where $\bar{u}$, $d$, $\nu_w$ are the mean velocity in the inlet, diameter of the inlet, and kinematic viscosity of water, respectively. Note that as the surface area of the mixing channel is equal to the sum of the inlet channel, the mean velocity of inlet and outlet coincides. The diffusion coefficient is fixed to $D=1.\bar{6}$ $10^{-9}\frac{m^2}{s}$, which yields a Schmidt number $Sc=\frac{\nu_w}{D}=600$ for water and $Sc=\frac{\nu_e}{D}\approx 450$ for ethanol. The Peclet number is  $Pe=\frac{\bar{u}d_h}{D}=Re\,Sc$. The diffusion coefficient of water-ethanol mixtures depends on the concentration \cite{legros2015investigation}, and this dependence should be considered in sub-grid models of diffusion controlled micromixing. Here we are interested in the macroscopic mixing, which is controlled by convection already at moderate Reynolds numbers ($Re>100$). We hence assume that the dependence of the diffusion coefficient on $\phi$ should have a minor impact on the macroscopic properties of mixing and so we neglect it. However, we note that it would be straightforward to incorporate a concentration-dependent diffusion coefficient $D(\phi)$ in our model and numerical code. 

\subsection{Boundary conditions}

The velocity profile at the inlet is taken as fully developed laminar flow in a square duct and can be obtained by solving the two-dimensional Poisson equation $\Delta u_1 = \frac{G}{\mu}$. The right hand side $\frac{G}{\mu}$ with pressure gradient $G$ is then scaled to the corresponding Reynolds number. The concentration is set to $\phi(x/d=-5.5)=0$ (water) and $\phi(x/d=5.5)=1$ (ethanol)  at the inflow planes. The no-slip boundary condition at the wall is employed for the velocity field, whereas a zero-flux boundary condition is used for the concentration. At the outlet plane a mass preserving convective boundary condition is used, defined as 
\begin{equation*}
 \frac{\partial \rho\boldsymbol{u}}{\partial t}+ U_\text{adv} \frac{\partial \rho\boldsymbol{u}}{\partial \boldsymbol{n}} = 0,
\end{equation*}
where the advective velocity $U_\text{adv}$ is set to be the mean velocity $\overline{u}$. Furthermore a zero flux boundary condition at the outlet plane is chosen for the concentration. A convective boundary condition was also tested and found not to influence the concentration field, while hindering convergence of our iterative solvers. 

\subsection{Diagnostic quantities}

We quantify the mixing efficiency with the degree of mixing
\begin{equation}
\delta = 1 - \frac{\sigma_b}{\sigma_\text{max}},
\label{eqn:dom}
\end{equation}
where $\sigma_\text{max}$ is the maximum of the variance achieved by completely segregated streams, and $\sigma_{b}$ is the mean deviation of the concentration profile as
\begin{equation}
\sigma^2_\text{max} = \overline{\phi}(1-\overline{\phi}) \text{ and } \sigma^2_{b} = \frac{1}{N} \sum_{i=1}^N (\phi_i-\overline{\phi})^2,
\end{equation}
where $\overline{\phi}$ is the mean concentration $\overline{\phi} = \frac{1}{N}\sum_{i=1}^N \phi_i$ of a cross section. The energy input is here quantified with the pressure loss
\begin{equation}
\Delta p = p_\text{out}-p_\text{in},
\label{eqn:pressLoss}
\end{equation}
where $p_\text{out}$ and $p_\text{in}$ are the surface-averaged pressure at the outlet and inlet cross section, respectively. Using the pressure loss $\Delta p$ the Darcy friction factor $\lambda$ is calculated as 
\begin{equation}
\lambda = 2\frac{\Delta p d_h}{L \bar{\rho} \bar{u}^2},
\label{eqn:darcy}
\end{equation}
where $L/d=18.5$ is the total length of the T-mixer and $d_h$ is the hydraulic diameter of the mixing outlet channel. Here $\bar{\rho} $ is chosen to be the mean density of the mixture. 

We characterise the relative importance of the acting transport mechanisms by considering three key time scales: the advective time scale $t_\text{adv} = \frac{d}{\bar{u}}$, the viscous time scale $t_\text{vis} = \frac{d^2}{\nu}$ and the diffusive time scale $t_\text{diff} = \frac{d^2}{D}$. Note that these can differ by several orders of magnitude depending on the flow state.

%%%%%%%%%%%%%%
% NUMERICS
%%%%%%%%%%%%%%
\section{Numerical discretization}
\label{sec:discretization}

Direct Numerical Simulations of turbulent flows must in principle resolve all flow scales in order to produce physically accurate results. In our case the largest eddies span the whole channel, whereas the smallest eddies are set by the Kolmogorov scale $\eta$. In practice a grid spacing $\Delta x$ of a few $\eta$ is required~\cite{pope,moinAnnual}, which results in very large numbers of grid points even at moderate Reynolds numbers. In addition, a statistical characterisation of the computed flow regimes implies that DNS need to span long time scales, in wall-bounded shear flows several advective time units ($t_\text{adv}$). Hence DNS are computationally very expensive: the required computing power scales as $Re^3$ under the assumption that algorithms scale optimally with problem size \cite{pope,moinAnnual}. This makes DNS prohibitive at large $Re$ and requires accurate numerical methods and their efficient implementation even for moderate $Re\in[10^3,10^4]$ as in LAS precipitation. 

\subsection{Spatial discretization}

We discretize the Navier--Stokes equations \eqref{eqn:tes}--\eqref{eqn:mcd} using a $2^\text{nd}$ order central finite-volume scheme in a collocated grid arrangement based on the work of Peric \emph{et al.}~\cite{Peric1988389}. For a detailed introduction to the finite volume method we refer the reader to the classic books of \citet{ferziger2012computational} and \citet{versteeg2007introduction}. The code used here (FASTEST-3D) has been developed at the Institute of Fluid Mechanics at the Friedrich-Alexander-Universit\"at Erlangen-N\"urnberg. It can solve for laminar and turbulent flows in complex geometries with DNS, LES and RANS. The code can be used also to solve fluid-structure \cite{breuer} and fluid-acoustic interaction \cite{schafer}, as well as chemical reactions \cite{enger2012czochralski}. FASTEST-3D is parallelized with a hybrid OpenMP-MPI implementation \cite{Scheit} and has been optimized for both vector computers with high workload per process and for massively parallel systems with relatively low workload per process but great communication overhead.

The spatial discretization of the convection equation \eqref{eqn:tem} is more challenging because of the large Schmidt number in our problem ($Sc=600$). Let us consider a WW mixture with a mean velocity of $\overline{u}=0.16 \frac{m}{s}$, corresponding to $Re=160$ in the inlet channels. At $Re=160$ the flow is stationary but in the engulfment regime, which is characterized by two counter-rotating vortices originating at the junction due to a Kevin-Helmholtz instability \citet{Fani20136}. Fig.~\ref{fig:WW160} shows that these vortices fold and stretch the contact plane between the two incoming fluid streams, resulting in sharp gradients in the concentration field $\phi$. The discretization of sharp solutions by central discretization schemes lead to wiggles, overshoots and numerical oscillations whereas pure upwind discretization schemes promote mixing by introducing numerical diffusion \cite{godunov1959difference}. Note that $Sc=600$  yields  $Pe = Re Sc = 96,000$, and in order to get smooth numerical approximations with a second order central scheme prohibitive grid spacings $\Delta \approx 10^{-8}m$ would be required. To alleviate this problem the total variation diminishing (TVD) criterion was developed to ensure monotonicity of the bounded variables in time and space \cite{harten1983high,yee1987construction,Waterson2007182}. Additionally, limiter functions are used with the help of locally calculated successive gradients, to add numerical diffusion only where sharp gradients occur and to preserve boundedness. The fundamental finite volume discretization of those can be found in \cite{versteeg2007introduction}.

Here we do not intend to present a new mathematical approach but rather show that TVD schemes perform remarkably well and are compulsory to maintain physical validity.  We performed a grid convergence study for the WW mixture at $Re=160$ in the engulfment regime shown. The relative error of the concentration field  \begin{equation}
 e_{\Delta}=\frac{||\phi_{\Delta}- \phi_{2 \Delta}||}{||\phi_{\Delta}||},
 \label{eqn:relerror}
 \end{equation}
shown in Fig.~\ref{fig:Scaling}(a), reflects an almost $2^{nd}$ order accuracy.  It is worth noting that sharp gradients are at about 45 degrees with respect to the control volume surfaces of our structured cartesian grid, corresponding to a worst case scenario.  

\subsection{Temporal discretization}

% I have changed this paragraph a bit, please check it carefully
DNS are often performed using semi-implicit schemes for advancing the Navier--Stokes equations in time \cite{moinAnnual}. We here employ an explicit method. Let us consider a WW mixture for $\overline{u}=0.3 \frac{m}{s}$ ($Re=300$ in the inlet channel), which corresponds to a chaotic flow state. In this case the flow field is well resolved with a grid spacing of $\Delta=\frac{1}{96} mm$. In order that explicit or semi-implicit temporal schemes be stable the Courant--Friedrichs--Lewy (CFL) condition must be satisfied, which requires in this case a time-step size of about $\Delta t_\text{adv} \approx C \Delta/(2 \overline{u}) \approx 1.7\cdot 10^{-5}s$ for a CFL constant of $C=0.5$. The molecular diffusion and viscous transport restrictions on stability for an explicit scheme are $\Delta t_\text{diff}=\Delta^2/(2D) \approx 3.25\cdot 10^{-2}s$ and $\Delta t_\text{vis}=\Delta^2/(2\nu) \approx 5.5\cdot 10^{-5}s$. We conclude that for $Re=300$ both fully explicit and semi-implicit schemes are stable provided that the CFL condition is satisfied and this remains so as the Reynolds number further increases. Hence we decided to perform our DNS with fully explicit scheme because of the much lower computing cost involved.

We chose a low-storage Runge-Kutta scheme decoupling equations \eqref{eqn:tes} and \eqref{eqn:mcd} from \eqref{eqn:tem}.  The right hand sides of \eqref{eqn:tes} and \eqref{eqn:tem} depend only on the spatial discretization, and for simplicity, these will be referred to as $Q_{\phi}$ and $Q_{\boldsymbol{u}}$, respectively. 
The time-stepping algorithm is given below. Here the subscripts $P$, $k$ refer to the central and surrounding center points of the control volumes respectively, as commonly used for finite volume discretizations \cite{Peric1988389,versteeg2007introduction}, and $\Delta V$ is the volume of the cell:
\begin{enumerate}[\itshape 1)]
\item A predictor step that solves the momentum equation with a 3-step Runge-Kutta time scheme
\begin{align}
\boldsymbol{u}_P^{\tau_1} &= \boldsymbol{u}_P^n+\alpha_1 \frac{\Delta t}{\rho^n \Delta V}\left( \boldsymbol{Q}_{\boldsymbol{u},P} + \sum_{k \in I_k}  \boldsymbol{Q}_{\boldsymbol{u},k} \right)^n \nonumber\\
\boldsymbol{u}_P^{\tau_2} &= \boldsymbol{u}_P^n+\alpha_2 \frac{\Delta t}{\rho^n \Delta V}\left( \boldsymbol{Q}_{\boldsymbol{u},P} + \sum_{k \in I_k}  \boldsymbol{Q}_{\boldsymbol{u},k}\right)^{\tau_1} \label{eqn:ruku}\\
\boldsymbol{u}_P^{*} &= \boldsymbol{u}_P^n+\alpha_3 \frac{\Delta t}{\rho^n \Delta V}\left( \boldsymbol{Q}_{\boldsymbol{u},P} + \sum_{k \in I_k}  \boldsymbol{Q}_{\boldsymbol{u},k} \right)^{\tau_2}, \nonumber
\label{eqn:ruku}
\end{align}
where $\boldsymbol{u_P}^*$ is not divergence free. 
\item Accounting for equation~\eqref{eqn:mcd} we formulate a corrector step for the pressure,  leading to a Poisson type equation 
\begin{equation}
\int_A \boldsymbol{n}  \cdot ( \nabla p') \, \mathrm{d}A =\frac{1}{\Delta t} \int_A \boldsymbol{n}  \cdot ( \rho \boldsymbol{u}^*) \, \mathrm{d}A.
\end{equation}
\item Solving the linear system for $p^{'}$
\item We obtain a new pressure field $p^{'}$ which complies the enforced Eq.~\ref{eqn:mcd} by updating the velocity field
\begin{equation}
\boldsymbol{u}_P^{**} = \boldsymbol{u}_P^{*} - \frac{\Delta t}{\rho_P} \nabla p^{'}|_P
\end{equation}
and the pressure field
\begin{equation}
p_P^{**} = p_P^{*}+p'. 
\end{equation} 
The corrector steps 2-4 are repeated until Eq.~\ref{eqn:mcd} is fulfilled within machine precision $\boldsymbol{u}_P^{n+1}=\boldsymbol{u}_P^{**}$ and $p_P^{n+1}=p_P^{**}$. 
\begin{itemize}
\item In the first correction loop $p^*=p^n$
\item If the stop criterion is not fulfilled $p^*=p^{**}$ and $u^*=u^{**}$
\end{itemize}
\item Eq.~\ref{eqn:tes} is solved with a 3-step Runge-Kutta time scheme and the new velocity field $\boldsymbol{u}^{n+1}$ 
\begin{align}
\phi_P^{\tau_1} &= \phi_P^n+\alpha_1 \frac{\Delta t}{\rho^n \Delta V}\left( Q_{\phi,P} + \sum_{k \in I_k}  Q_{\phi,k}\right)^n \\
\phi_P^{\tau_2} &= \phi_P^n+\alpha_2 \frac{\Delta t}{\rho^n \Delta V}\left( Q_{\phi,P} + \sum_{k \in I_k}  Q_{\phi,k} \right)^{\tau_1} \\
\phi_P^{n+1} &= \phi_P^n+\alpha_3 \frac{\Delta t}{\rho^n \Delta V}\left( Q_{\phi,P} + \sum_{k \in I_k}  Q_{\phi,k} \right)^{\tau_2}. 
\end{align}
\item update viscosity $\mu$ and density $\rho$ with $\phi^{n+1}$
\end{enumerate}

\subsection{Grid study}

We determined the required spatial resolution by two criteria. Firstly, we enforce that the difference in the pressure drop \eqref{eqn:pressLoss} computed with the chosen grid and a coarser grid is less than 1\%. Secondly, in the mixing zone, where strong vortical structures arise, the grid spacing $\Delta$ should be in the order of a few Kolmogorov length scales $\eta$. In fact, we find that for the turbulent flow states $\frac{\Delta}{\eta} \approx 1.2 \pm 0.7$ suffices. The  settings for the simulations in this paper are summarized in Table~\ref{tab:resolution}. 

\begin{table}
\begin{center}
\begin{tabular}{cccccc}
 \\ \hline\toprule
$Re$ & mixture & $\Delta x$,  $\Delta y$,  $\Delta z$ in [mm]  & \#grid points & $\Delta t$ in [$10^{-6}$s] & $t/t_\text{adv} $\\ \hline \midrule 
220--650 & WW & $\frac{1}{96}$, $\frac{1.5}{96}$,$\frac{1}{96}$& $23\cdot 10^6$ &3.5--1& 30--50\\ \midrule
300-650 & WE & $\frac{1}{96}$, $\frac{1.5}{96}$,$\frac{1}{96}$&$23\cdot 10^6$ & 2--1& 30--50\\ \midrule
1100 & WW,WE & $\frac{1}{128}$, $\frac{1.5}{128}$,$\frac{1}{128}$&$55\cdot 10^6$ & 0.4 & 40-60\\ \midrule
2000 & WW,WE & $\frac{1}{160}$, $\frac{1.5}{160}$,$\frac{1}{160}$&$80\cdot 10^6$ & 0.08 & 40-60\\  \bottomrule \hline 
\end{tabular}
\caption{Overview of simulation settings. Grid spacings $\Delta$ are given for the mixing channel. In the junction an equidistant grid spacing is chosen having the finer shown resolution. Time steps $\Delta t$ are selected such that more than 80\% of the simulation time the CFL condition is satisfied with $C<0.3$ everywhere in the domain. Depending on the branch and initial guess it takes up to 60 advective time units to reach statistical steady state (last column).}
\label{tab:resolution}
\end{center}
\end{table}

\subsection{Code performance}

\begin{figure}[h!]
  \begin{center}
    \begin{tabular}{cc}
      (a) & (b)\\
      \includegraphics[width=0.4\linewidth]{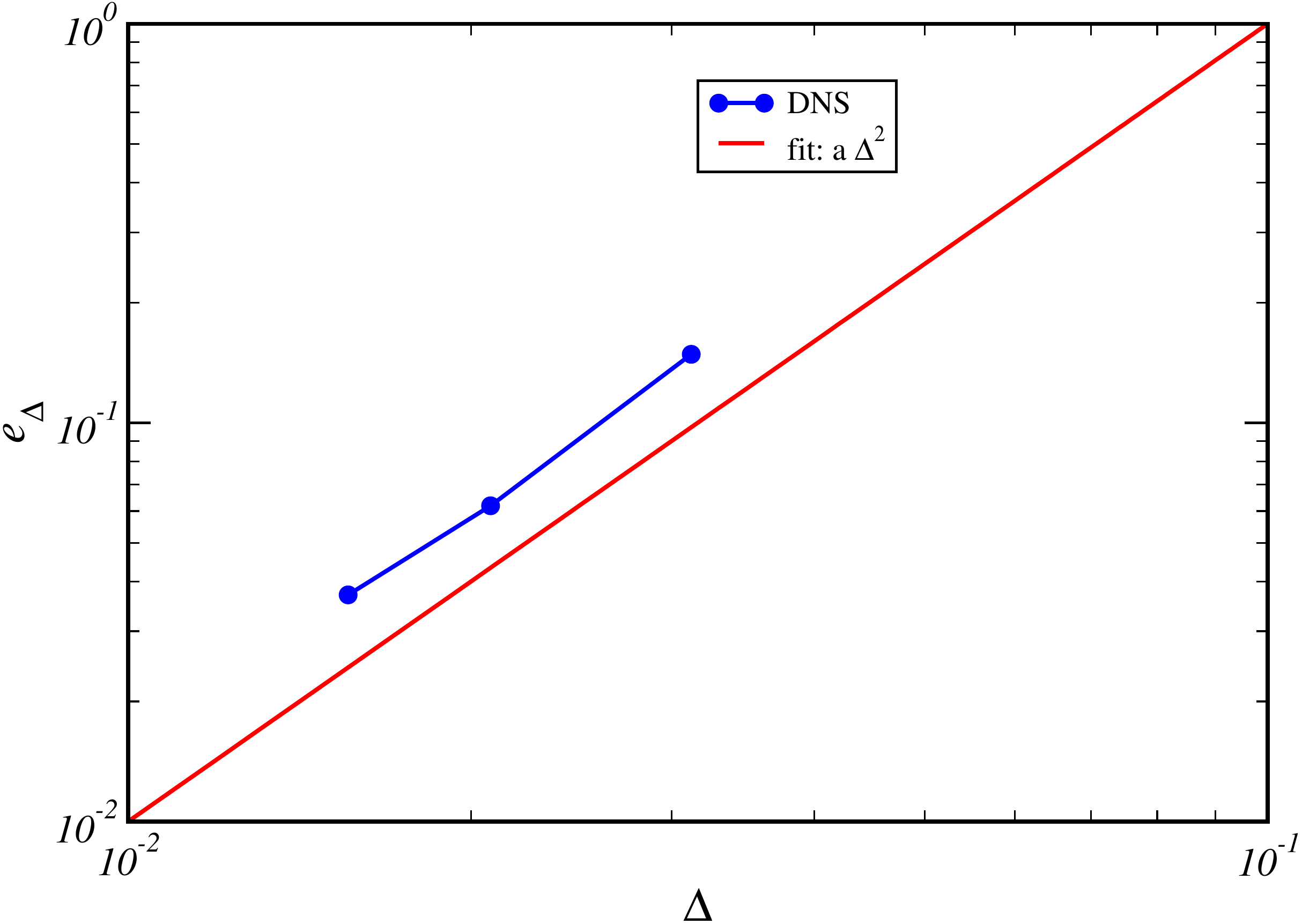} &
      \includegraphics[width=0.4\linewidth]{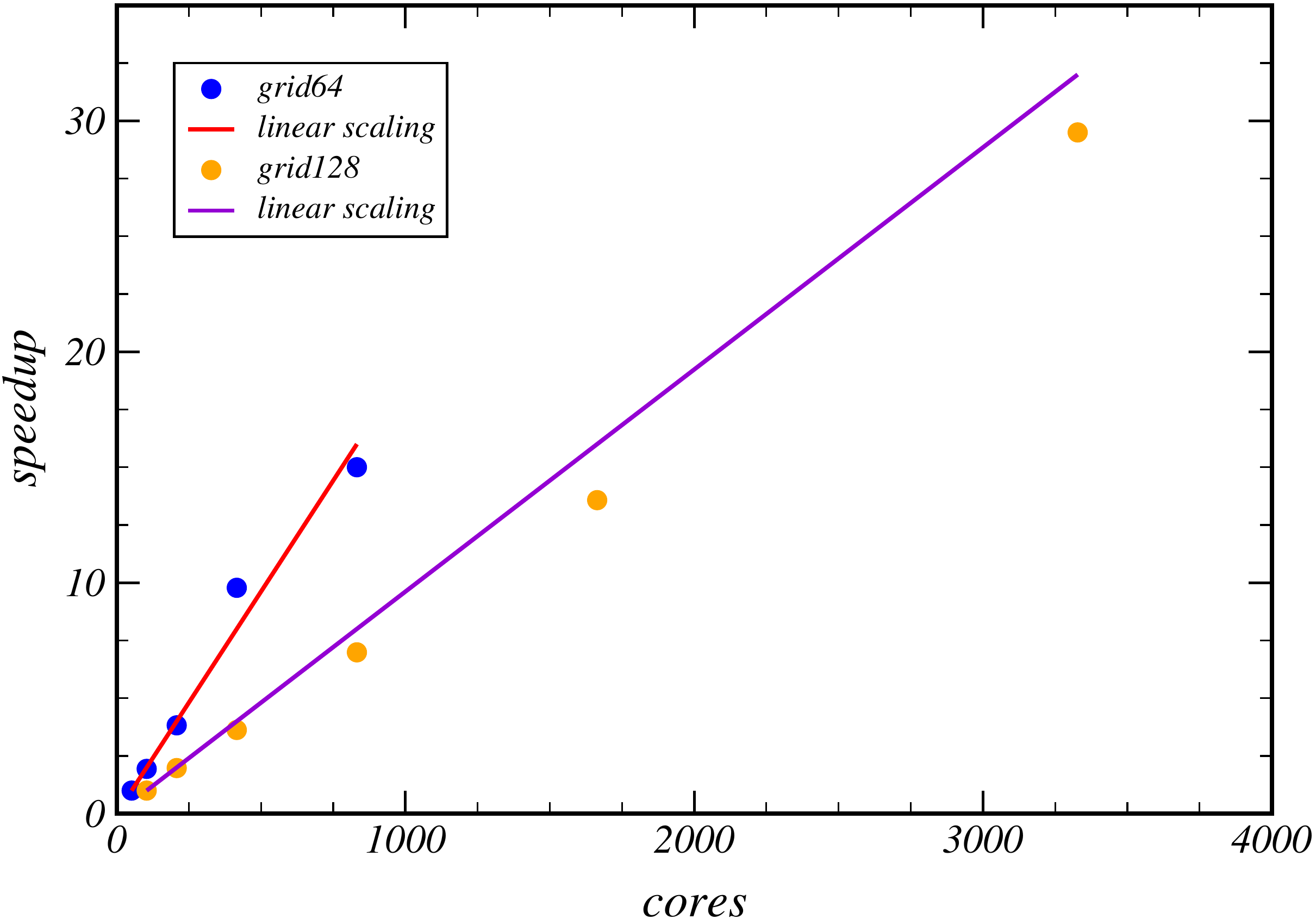}
    \end{tabular}
        
  \end{center}
  \caption{(a) Relative error $e_{\Delta}$ of the concentration field \eqref{eqn:relerror} as function of the grid spacing $\Delta$. The TVD scheme with a \textit{van Leer} limiter function exhibits spatial convergence of approximately seocod order $e_{\Delta}=\mathcal{O}(\Delta^{2})$. (b) Speed-up $\frac{T}{T_1}$ (strong scaling) over cores for two different grids sizes. Lines show perfect scaling for the corresponding setup.}
\label{fig:Scaling}
\end{figure}

Simulating turbulent flow at $Re \approx \mathcal{O}(10^3)$ requires the numerical code to scale in a supercomputer. The  performance of our code is demonstrated in~Fig.~\ref{fig:Scaling} with a speedup calculation for two sample setups grid64 (7 Mio. control volumes) and grid128 (55 Mio. control volumes). The speedup is calculated by $\frac{T}{T_1}$, where $T$ is the mean execution time for one time step and $T_1$ is the mean execution time for the smallest number of cores. Ten sample time steps for a fixed number of pressure corrections is used to calculate the mean execution time. Notably, the large setup loses only 5\% performance up to 3000 processors. A plain MPI-parallelization strategy is adopted for the implementation of the code. The scaling test has been realized on SuperMUC at the Leibniz Supercomputing Centre (LRZ) in  Germany.

%%%%%%%%%%%%%%
% Benchmark
%%%%%%%%%%%%%%
\section{Benchmark}
\label{sec:benchmark}

We have tested our code for a wide range of Reynolds numbers $Re \in [10,2000]$ and different geometries, including classical test cases for DNS such as channel flow \cite{moin}. In this section we present two test cases of mixing of miscible fluids in a T-junction. We provide a quantitative comparison against published results and discuss the physics of mixing as a reference for the cases at higher Reynolds number presented in~\S\ref{sec:transToTurb}.

\subsection{Mixing of water-water in a time-periodic regime}

\begin{figure}
  \begin{center}
    \begin{tabular}{ccc}
      (a) \small{$y/d=-0.25$} & (b) \small{$y/d=-5.5$}& (c) \small{$y/d=-12$}\\
      \includegraphics[width=0.25\linewidth]{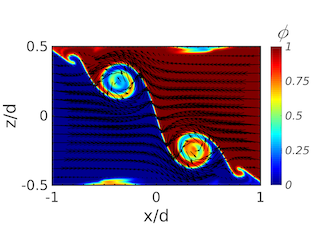} &
      \includegraphics[width=0.25\linewidth]{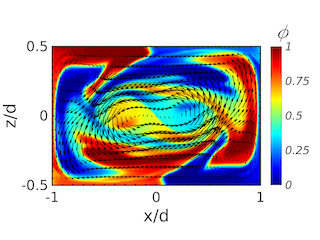} &
      \includegraphics[width=0.25\linewidth]{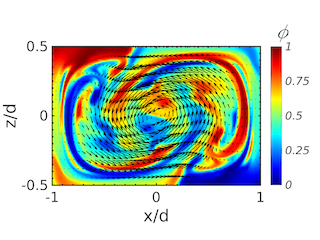} \\
      \includegraphics[width=0.25\linewidth]{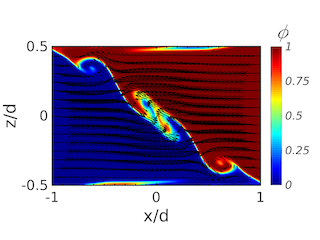} &
      \includegraphics[width=0.25\linewidth]{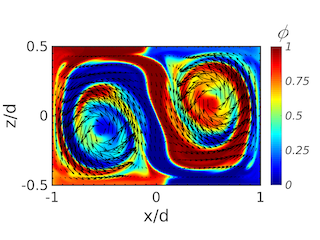} &
      \includegraphics[width=0.25\linewidth]{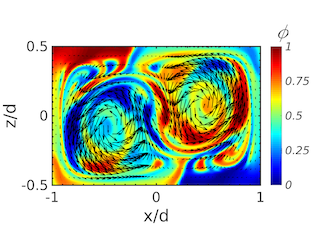} 
	\end{tabular}
   \end{center}
\caption{Snapshots of the colormap of the concentration for WW in the time-periodic engulfment regime ($Re=220$) at $t/t_{adv} \approx 3$ (top row) and $t/t_{adv} \approx 5$ (bottom row) corresponding to the two different phases of mixing as it can be seen in Fig.~\ref{fig:Re220WWdom}. The arrows show the cross-sectional velocity field.}
  \label{fig:Re220WWsnaps}
\end{figure}

\begin{figure}
   \begin{center}
       \begin{tabular}{cc}
                      (a) & (b) \\
           \includegraphics[width=0.4\linewidth]{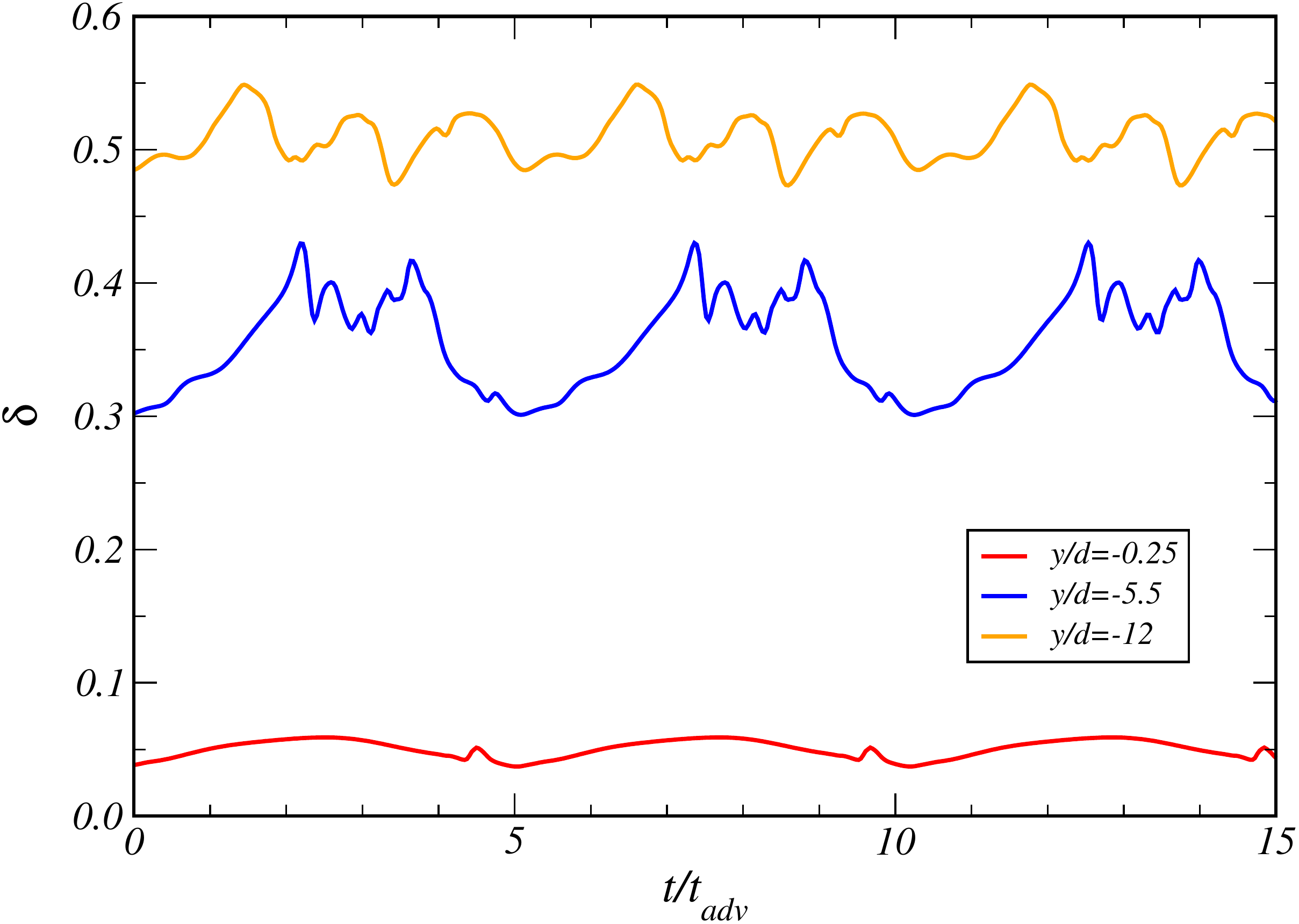}   &
           \includegraphics[width=0.4\linewidth]{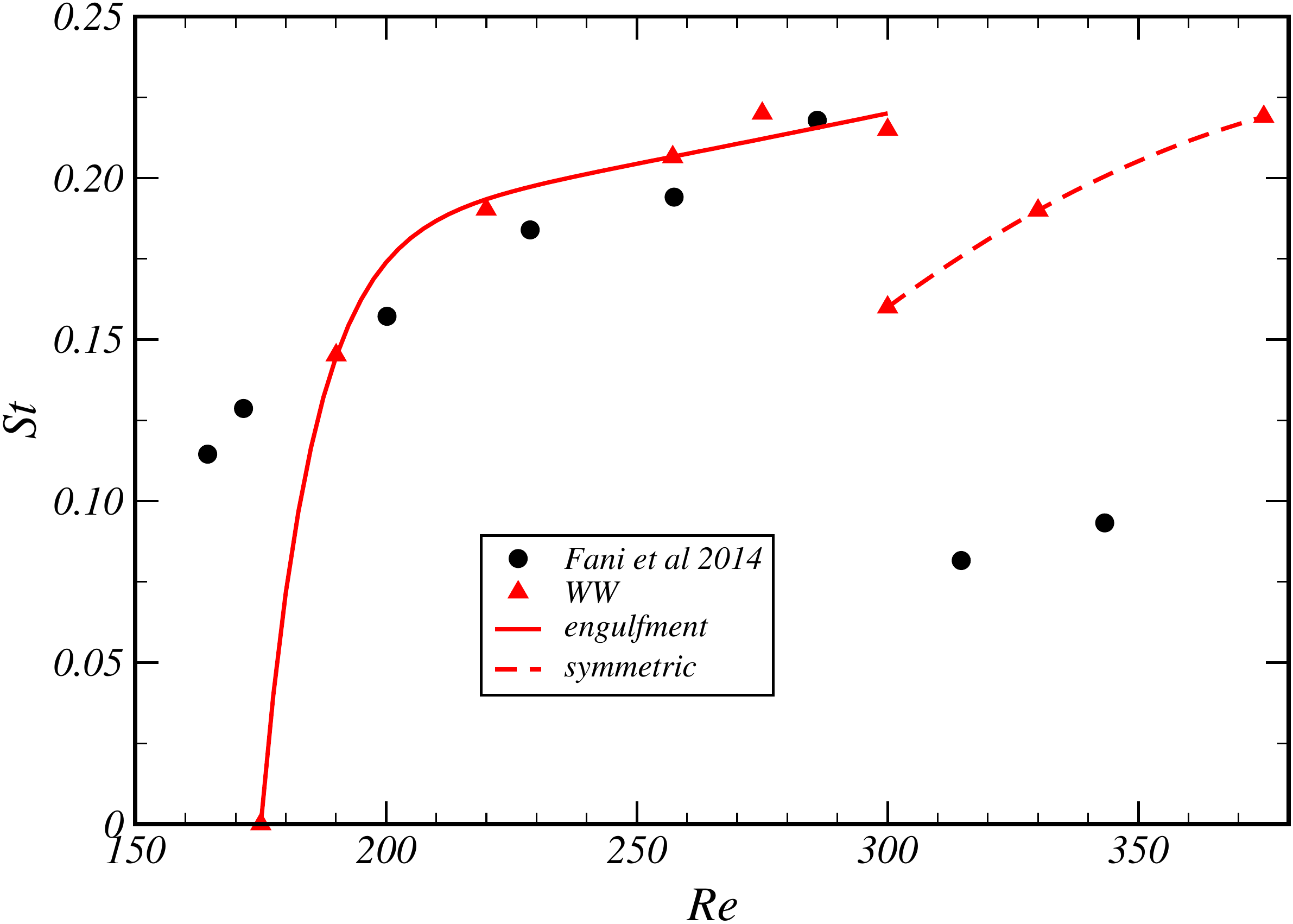}
       \end{tabular}      
  \end{center}
  \caption{(a) Time-evolution of the mixing degree $\delta$ at three selected cross sections for $Re=220$  in the time-periodic engulfment regime. The time-averaged mixing degree is $\delta =(0.048,0.39,0.51)$, respectively. (b) Strouhal number as a function of Reynolds number (red circles, the line is to guide the eyes). The data from the simulations of \citet{Fani20147} is shown as black triangles. Because of the slightly different geometry their data have been scaled here with the inlet hydraulic diameter to allow for comparison.}
  \label{fig:Re220WWdom}
\end{figure}  

\citet{thomas2010experimental} and \citet{Fani20147} showed experimentally and numerically that as the Reynolds number increases the steady engulfment regime shown in~Fig.~\ref{fig:WW160} is superseded by an unsteady periodically oscillating regime. The intrinsic Kevin-Helmholtz vortex structures persist, but, in comparison to the steady state, the vortex core moves with time as shown in the two snapshots of Fig.~\ref{fig:Re220WWsnaps}(a). Note also that the vortices are more intense and convoluted, and hence the spatially-averaged mixing degree $\delta$ at $y=-5.5$ and $-12$ has more than doubled from $Re=160$ to $Re=220$. The time-periodic nature of this regime can be seen in~Fig.~\ref{fig:Re220WWdom}(a), which shows time-series of $\delta$ for the three cross-sections in~Fig.~\ref{fig:Re220WWsnaps}. The frequency of the oscillation $f$ , expressed with the dimensionless Strouhal number $St=\frac{fd}{\bar{u}}$, is shown in~Fig.~\ref{fig:Re220WWdom}(b) as a function of Reynolds number. Here the onset of unsteadiness occurs at $Re\approx 175$ and our simulations are in very good agreement with those of \citet{Fani20147} in spite of the different width-to-height aspect-ratio of the inlets (0.75  in \cite{Fani20147} and 1 here), outlet (1.5 in \cite{Fani20147} and 2 here) and length of the outlet channel (25 in \cite{Fani20147} and 12.5 here). Surprisingly, the frequency of the natural flow pulsation is rather insensitive to the details of the geometry. Note that at  $Re \approx 275$ there is a sharp change in the oscillation frequency, which was attributed in \cite{thomas2010experimental,Fani20147} to a transition to a different more symmetric time-periodic flow regime. As this is also observed here, it appears that this is also a robust feature of WW flows at T-junctions. 
 
\subsection{Water-ethanol mixing in the steady engulfment regime}

\begin{figure}[h!]
  \begin{center}
    \begin{tabular}{ccc}
          (a) \small{$y/d=-0.25$} & (b) \small{$y/d=-5.5$}& (c) \small{$y/d=-12$}\\
      \includegraphics[width=0.25\linewidth]{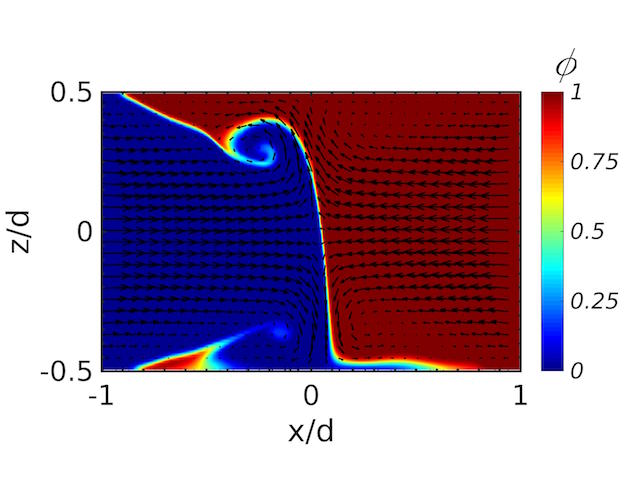} &
      \includegraphics[width=0.25\linewidth]{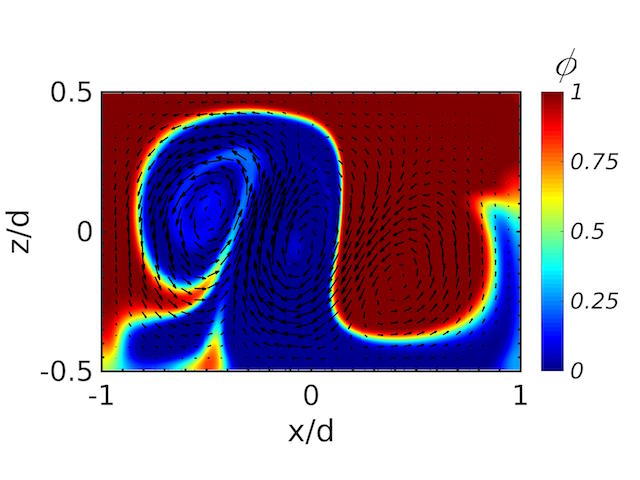} &
      \includegraphics[width=0.25\linewidth]{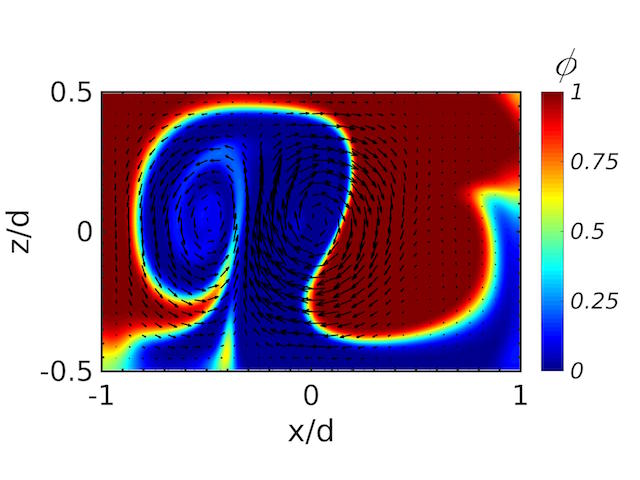} \\
      \includegraphics[width=0.25\linewidth]{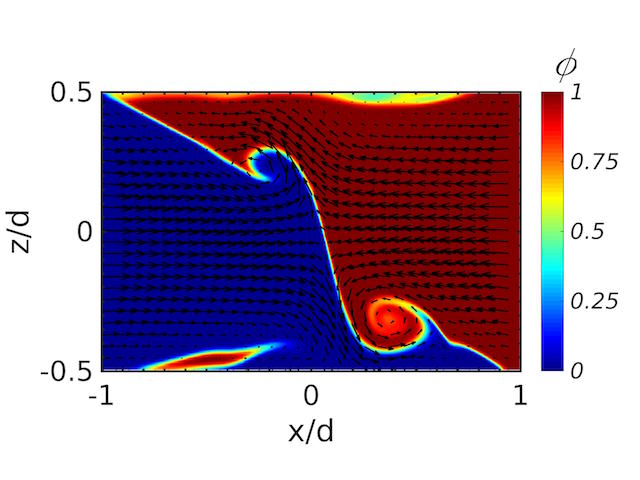} &
      \includegraphics[width=0.25\linewidth]{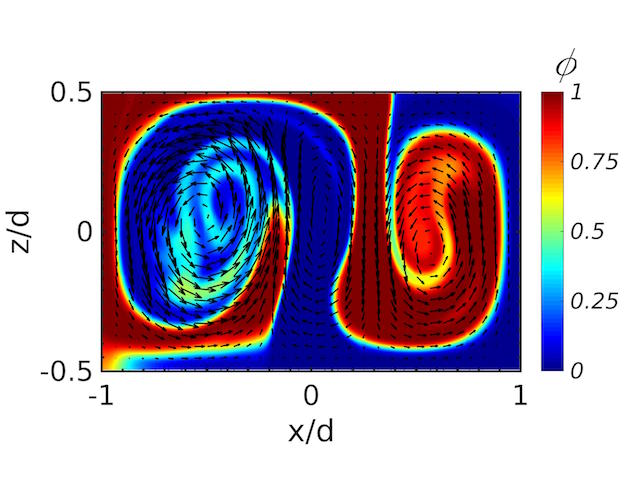} &
      \includegraphics[width=0.25\linewidth]{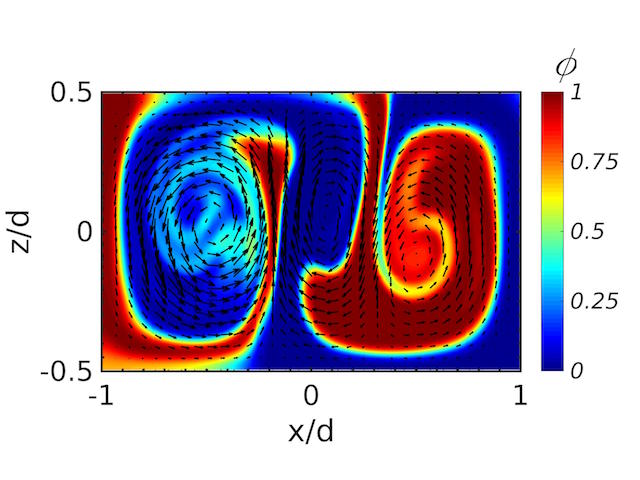} 
       \end{tabular}
  \end{center}
\caption{Colormaps of the concentration for WE in the steady engulfment regime at $Re=160$ (top row) and $Re=220$ (bottom row). The arrows show the cross-sectional velocity field. 
}
  \label{fig:Re160220WE}
\end{figure}

We validated our code for WE mixtures by performing simulations of the steady engulfment regime at $Re=160$ and $Re=220$. The cross-sectional profiles of the concentration shown in~Fig.~\ref{fig:Re160220WE} are in excellent agreement with those of \citet{Orsi2013174} (see their Figure~6c--d). Interestingly, the emergence of the vortices near the junction as seen in~Fig.~\ref{fig:Re160220WE}(a) is comparable to that of the WW case as shown in~Fig.~\ref{fig:WW160}(a). \citet{Orsi2013174} found that in  comparison to WW, in WE the transition to the engulfment regime is delayed to larger $Re$ and that the flow remains steady up to larger $Re$. This is because of the stabilizing effect resulting from the increased viscosity of the mixture (see~Fig.~\ref{fig:TJunctionPFit}b). Thus, a comparison of the flow features at $Re=160$ (WW) and $Re=220$ (WE) is appropriate. In particular the mixing degree at $y/d = (-0.25, -5.5, -12)$ is found to be in the same order as $\delta =(0.041,0.219,0.243)$ for WW and $\delta =(0.038,0.174,0.21)$ for WE, respectively. Note however that in WE the nonlinear dependence of the viscosity on the concentration makes the vortices asymmetric. In particular, the different strength of the vortices at their origin ($y/d=0$) generates a completely asymmetric flow pattern along the mixing channel. 

%%%%%%%%%%%%%%
% TRANSITION
%%%%%%%%%%%%%%
\section{Transition to turbulence and multiplicity of flow states}
\label{sec:transToTurb} 

\begin{figure}
  \begin{center}
  \begin{tabular}{cc} 
      (a) & (b)\\
      \includegraphics[width=0.4\linewidth]{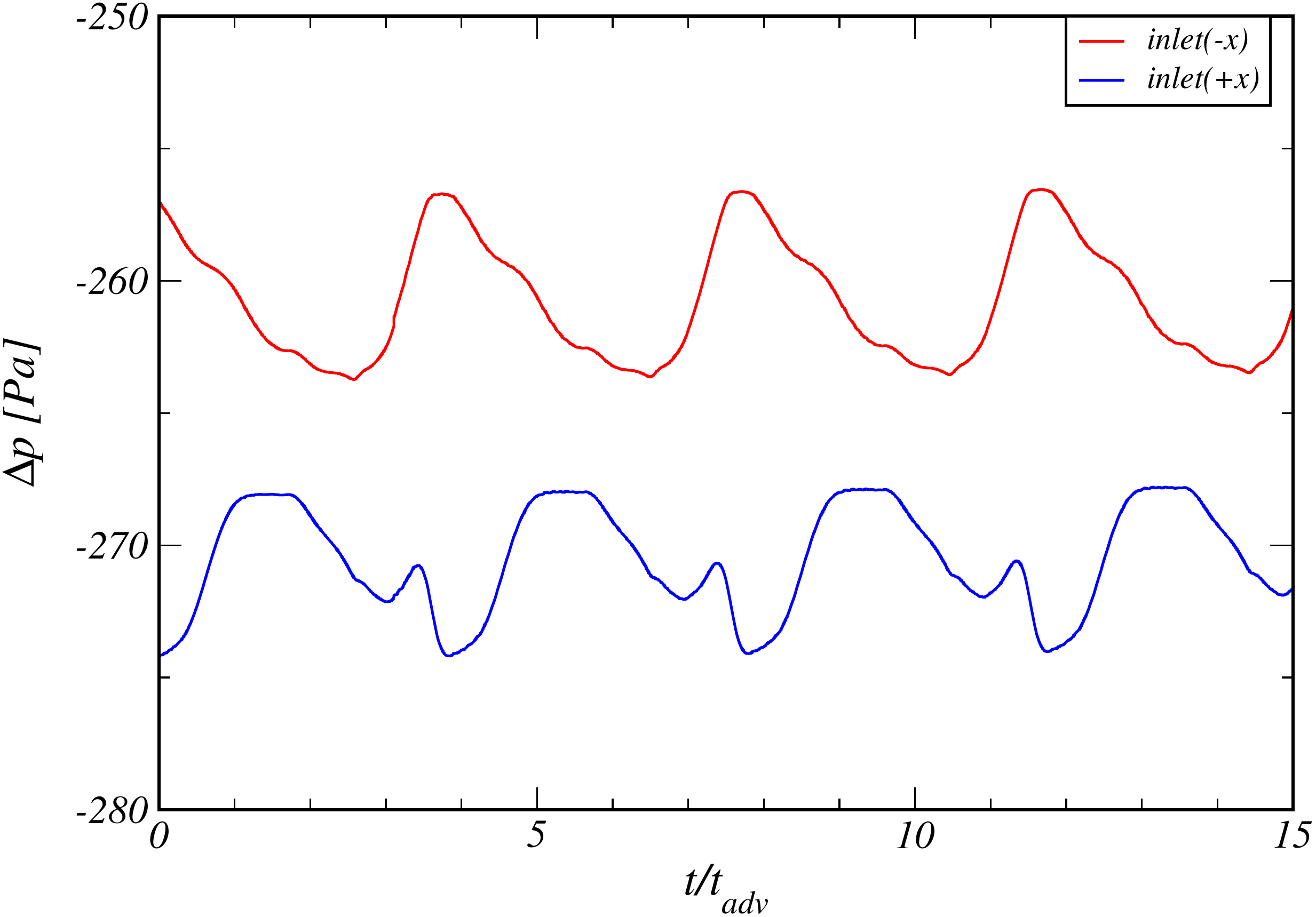} &
      	\includegraphics[width=0.4\linewidth]{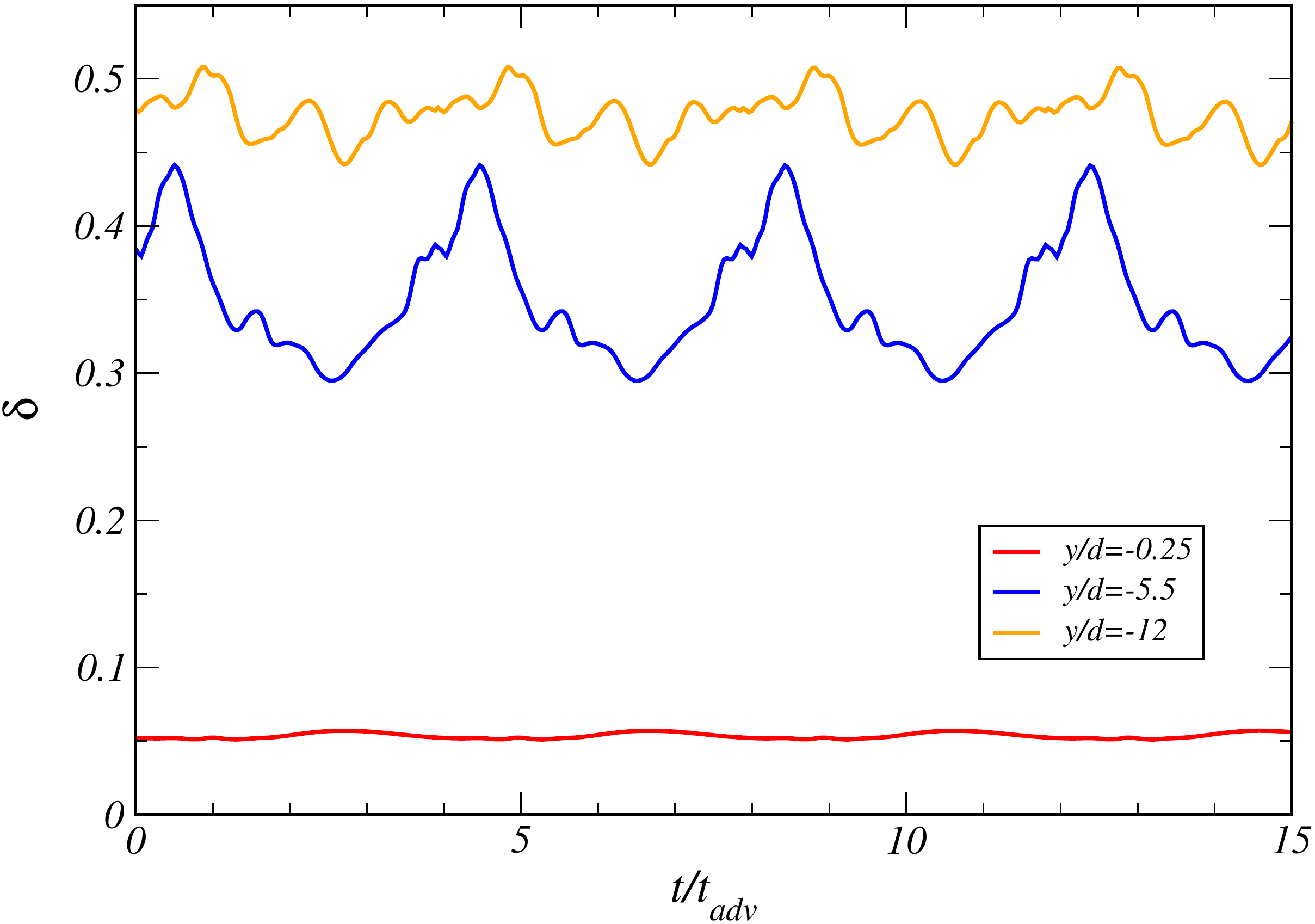}
      \end{tabular}
  \end{center}
  \caption{Time-periodic engulfment regime for WE at $Re=300$. (a) Time evolution of surface averaged pressure loss $\Delta p$ computed for water ($\phi=0$; $x/d=-5$) and ethanol ($\phi=1$; $x/d=5$) inlets. (b) Time evolution of the mixing degree $\delta$ at three selected cross sections (see legend). The time-averaged mixing degree is $\delta =(0.048,0.35,0.47)$, respectively. }
  \label{fig:WE300PressLoss}
\end{figure}

We investigated the transition to turbulence of WE mixtures by initializing our simulations from the steady engulfment regime at $Re=160$ and then stepwise changing $Re$ to a prescribed value. We found that for $Re > 225$ the flow starts pulsating periodically after a long transient of about $t/t_\text{adv} > 100$. The resulting time-series of the pressure loss necessary to drive the flow at the left (-x) and right (+x) inlet are shown in Figure~\ref{fig:WE300PressLoss}(a) for $Re=300$. As ethanol is more viscous than water, a higher pressure difference is necessary to drive ethanol at the same volumetric flow rate as water. Figure~\ref{fig:WE300PressLoss}(b) shows the time evolution of the spatially-averaged mixing degree $\delta$ at three selected cross-sections. As in the WW case shown in Fig.~\ref{fig:Re220WWdom}, a quiescent phase with relatively poor mixing is followed by a more dynamical phase characterized by a sharp peak in the mixing degree. By averaging these time series over a period we find that the mixing degree at $Re=300$ is more than twice that of the steady engulfment regime at $Re=220$. Overall, our results are in qualitative agreement with the experiments of \citet{Wang2012252}, who reported a strong increase in the mixing efficiency at $Re\approx 200$ for WE mixtures. Note however that their experiments were carried out in a T-junction with the same dimensions for the inlet and outlet channels, which prevents a direct comparison of our simulations and their experiments. 

\begin{figure}
  \begin{center}
    \begin{tabular}{ccc}
    (a) \small{$y/d=-0.25$} & (b) \small{$y/d=-5.5$}& (c) \small{$y/d=-12$} \\
      \includegraphics[width=0.25\linewidth]{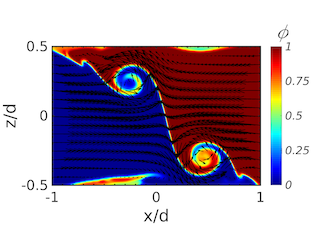} &
      \includegraphics[width=0.25\linewidth]{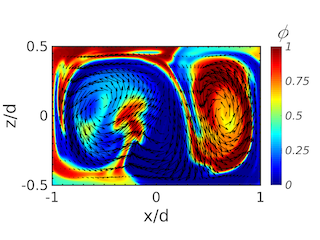} &
      \includegraphics[width=0.25\linewidth]{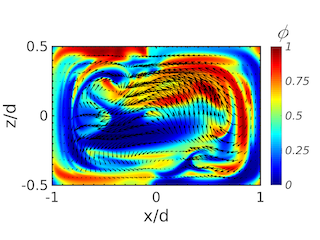} \\
      \includegraphics[width=0.25\linewidth]{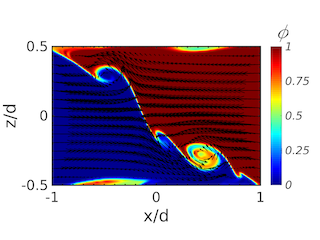} &
      \includegraphics[width=0.25\linewidth]{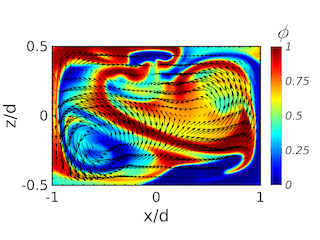} &
      \includegraphics[width=0.25\linewidth]{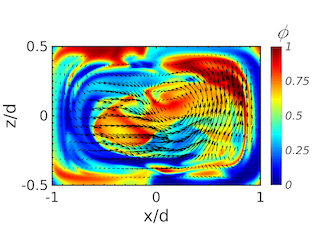}\\
      \includegraphics[width=0.25\linewidth]{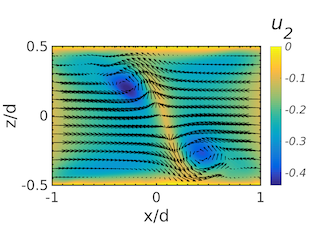} &
      \includegraphics[width=0.25\linewidth]{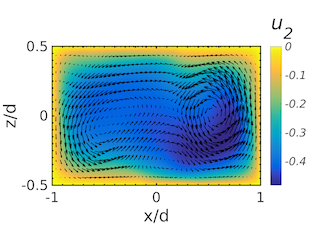} &
      \includegraphics[width=0.25\linewidth]{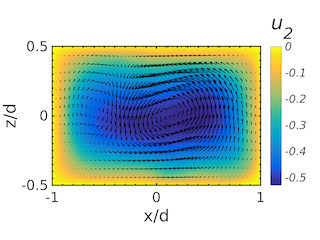}  
	\end{tabular}
   \end{center}
\caption{Time-periodic engulfment regime for WE at $Re=300$. The top and middle row show snapshots of the concentration (colormap) and cross-sectional velocity field (arrows) at $t/t_{adv} \approx 2.5$ and $t/t_{adv} \approx 4.5$, respectively. The bottom row shows the time-averaged streamwise (colormap) and  cross-sectional velocities (arrows).}
\label{fig:WE300snaps}
\end{figure}

Overall the WE time-periodic engulfment regime is qualitatively similar to that in WW reported by \citet{Fani20136} and described in \S\ref{sec:benchmark} for our geometry. However, the asymmetry in the viscosity as a function of concentration leads to different shear stresses at the contact plane between the two fluids and hence the arising Kelvin-Helmholtz vortices shown in Fig.~\ref{fig:WE300snaps}(a) are slightly asymmetric. The growth and the intense convolution of the vortices substantially enhances mixing along the main channel, as can be seen in the snapshots of Fig.~\ref{fig:WE300snaps}(b)--(c). This increase in mixing carries an increase of the mixture's viscosity and hence towards the outlet vortices are damped and the flow starts to relaminarize. This process can be seen in the time-averaged streamwise velocity at the same cross sections (bottom row of Fig.~\ref{fig:WE300snaps}). Near the junction ($y/d=-0.25$) an almost rotationally symmetric flow profile, which is a landmark of the engulfment state, is observed and progressing towards the outlet the irregular pattern of $y/d=-5.5$ slowly evolves into a quasi-parabolic flow profile at $y/d=-12$. 

\begin{figure}
  \begin{center}
    \begin{tabular}{ccc}
   (a) \small{$y/d=-0.25$} & (b) \small{$y/d=-5.5$}& (c) \small{$y/d=-12$} \\
      \includegraphics[width=0.25\linewidth]{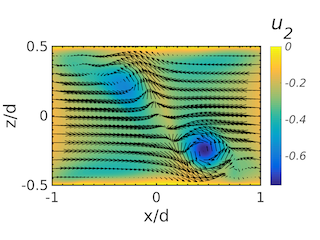} &
      \includegraphics[width=0.25\linewidth]{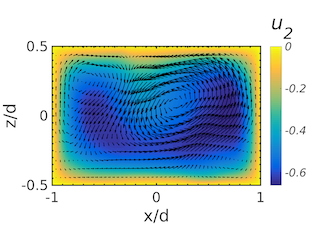} &
      \includegraphics[width=0.25\linewidth]{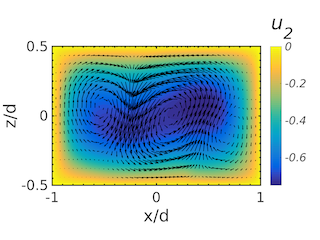} \\
      \includegraphics[width=0.25\linewidth]{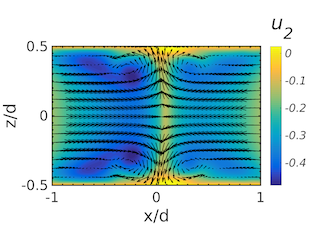} &
      \includegraphics[width=0.25\linewidth]{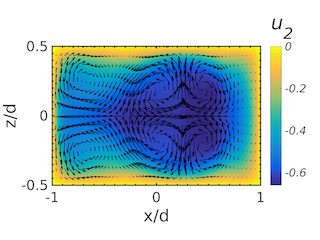} &
      \includegraphics[width=0.25\linewidth]{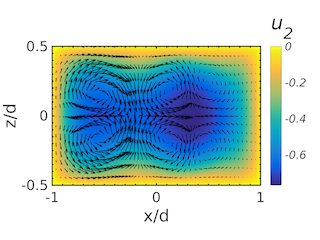} 
	\end{tabular}
   \end{center}
\caption{Colormap of the time-averaged streamwise velocity for WE at $Re=425$ in the chaotic engulfment regime (top row) and time-periodic symmetric regime (bottom row). The arrows show the time-averaged cross-sectional velocity field. }
\label{fig:WE425snaps}
\end{figure}

\begin{figure}
  \begin{center}
    \begin{tabular}{cc}
     (a) & (b) \\
      \includegraphics[width=0.415\textwidth]{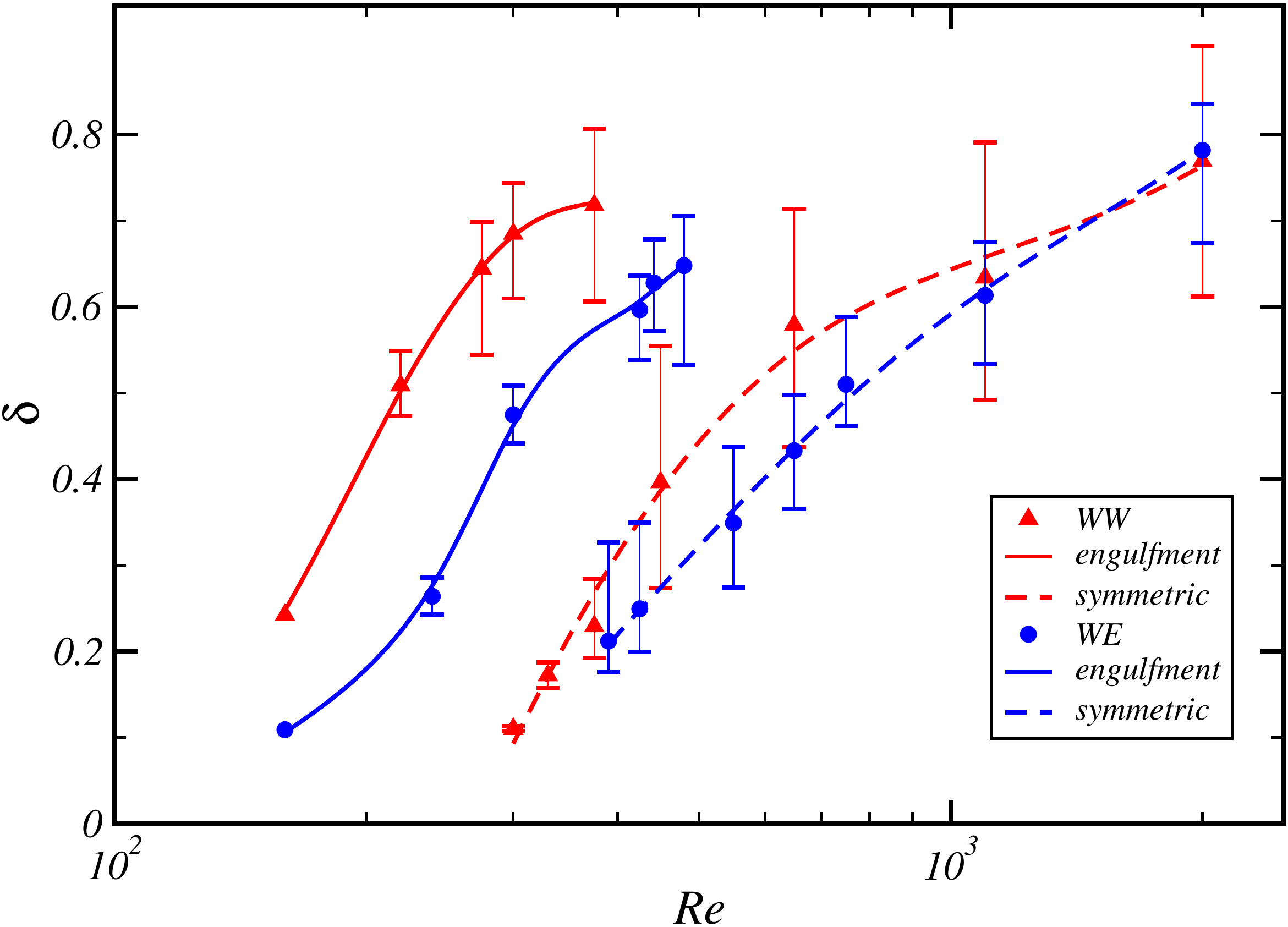} &
  		\includegraphics[width=0.4\textwidth]{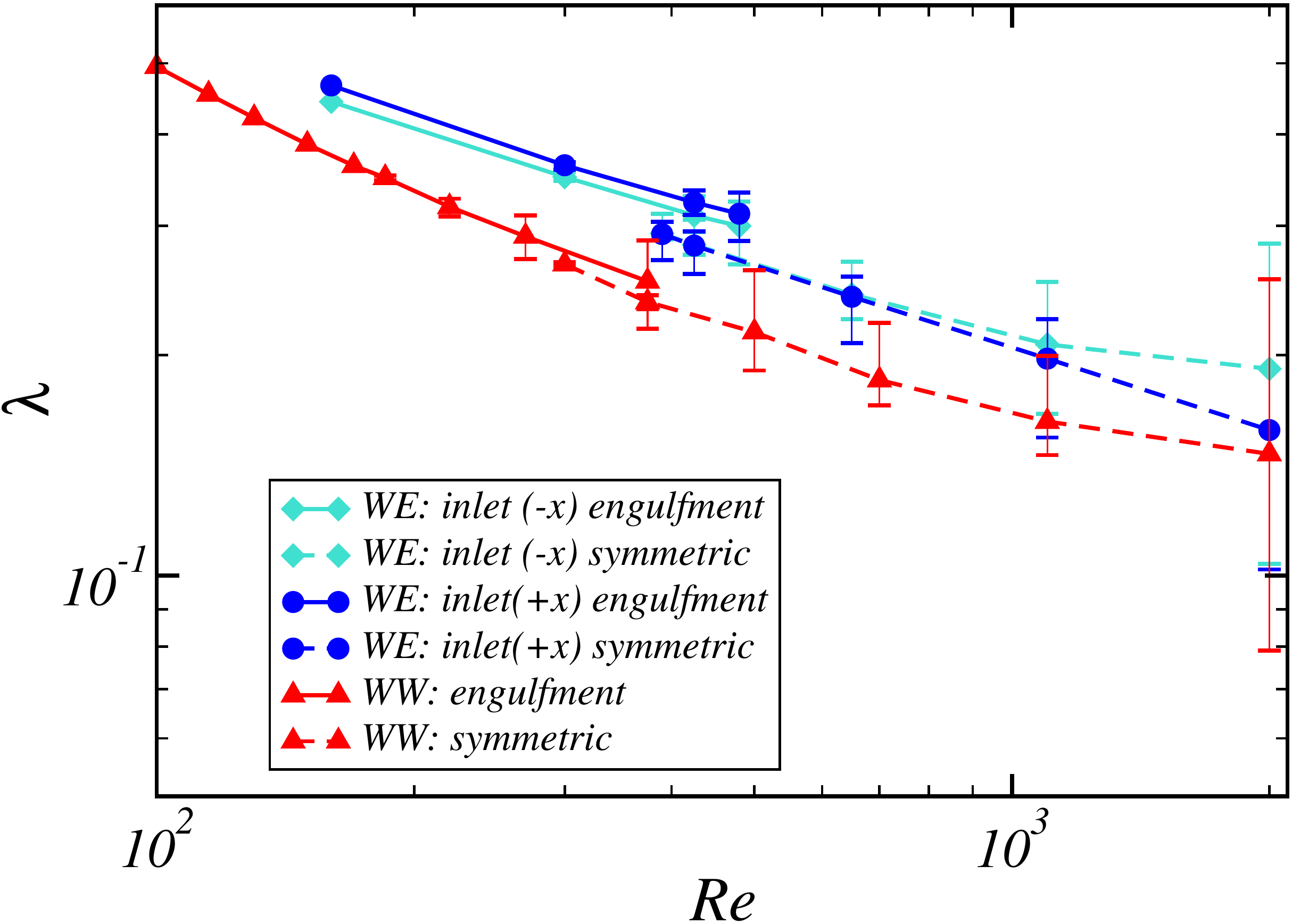} 
    \end{tabular}
  \end{center}
	\caption{(a) Mixing degree $\delta$ evaluated at cross section $y/d=-12$ as a function of Reynolds number. (b) Darcy friction factor $\lambda$ as a function of Reynolds number in a log-log scale. Error bars illustrate the maximum and minimum values of $\delta$ and	$\lambda$ for an unsteady flow, respectively.}
	\label{fig:DarcyFrictionFactor}
\end{figure}

As the Reynolds number is further increased the flow becomes first quasi-periodic and finally chaotic. The top panel of Fig.~\ref{fig:WE425snaps}(a) shows that in average an approximate rotational symmetry is retained near the junction, and hence we term this as the chaotic engulfment regime. The onset of chaotic mixing continues to enhances the performance of the T-mixer, until at $Re \approx 500$ the mixing degree suddenly drops by a factor of three (see  Fig.~\ref{fig:DarcyFrictionFactor}a). This dramatic change in the mixing behavior is caused by the transition to a new top-down symmetric time-periodic flow state (see the bottom row of  Fig.~\ref{fig:WE425snaps}). We investigated this surprising phenomenon by subsequently reducing the Reynolds number and found a hysteresis region in which the chaotic engulfment regime and the symmetric regime coexist. Note in fact the two flow states shown in Fig.~\ref{fig:WE425snaps} were both obtained at $Re=425$. Interestingly, Figure~\ref{fig:DarcyFrictionFactor}(b) shows that the pressure drop to drive the flow at a constant mass flux at a given number is higher for the engulfment regime than for the symmetric regime.

In the symmetric regime the water and ethanol streams flow in a nearly parallel fashion, thus strongly hindering mixing. Although there are vortices deforming the interface between the fluids locally, the intertwining of the two streams characteristic of the engulfment regime is entirely missing.  By subsequently increasing $Re$, the symmetric flow suffers a transition to turbulence. As a result the mixing quality starts to slowly recover, however, it is only at $Re \approx1100$ that the same mixing degree as at $Re \approx 500$ in the engulfment regime is recovered. We note that the transition to turbulence and evolution of the mixing as a function of Reynolds number are similar for WW (see Fig.~\ref{fig:DarcyFrictionFactor}a) and hence not discussed here in detail. 

%%%%%%%%%%%%%%
% TURBULENCE
%%%%%%%%%%%%%%
\section{Turbulent regime}

\begin{figure}
  \begin{center}
  	\begin{tabular}{cc}
  	(a) WE & (b) WW \\
  	  \includegraphics[width=0.4\textwidth]{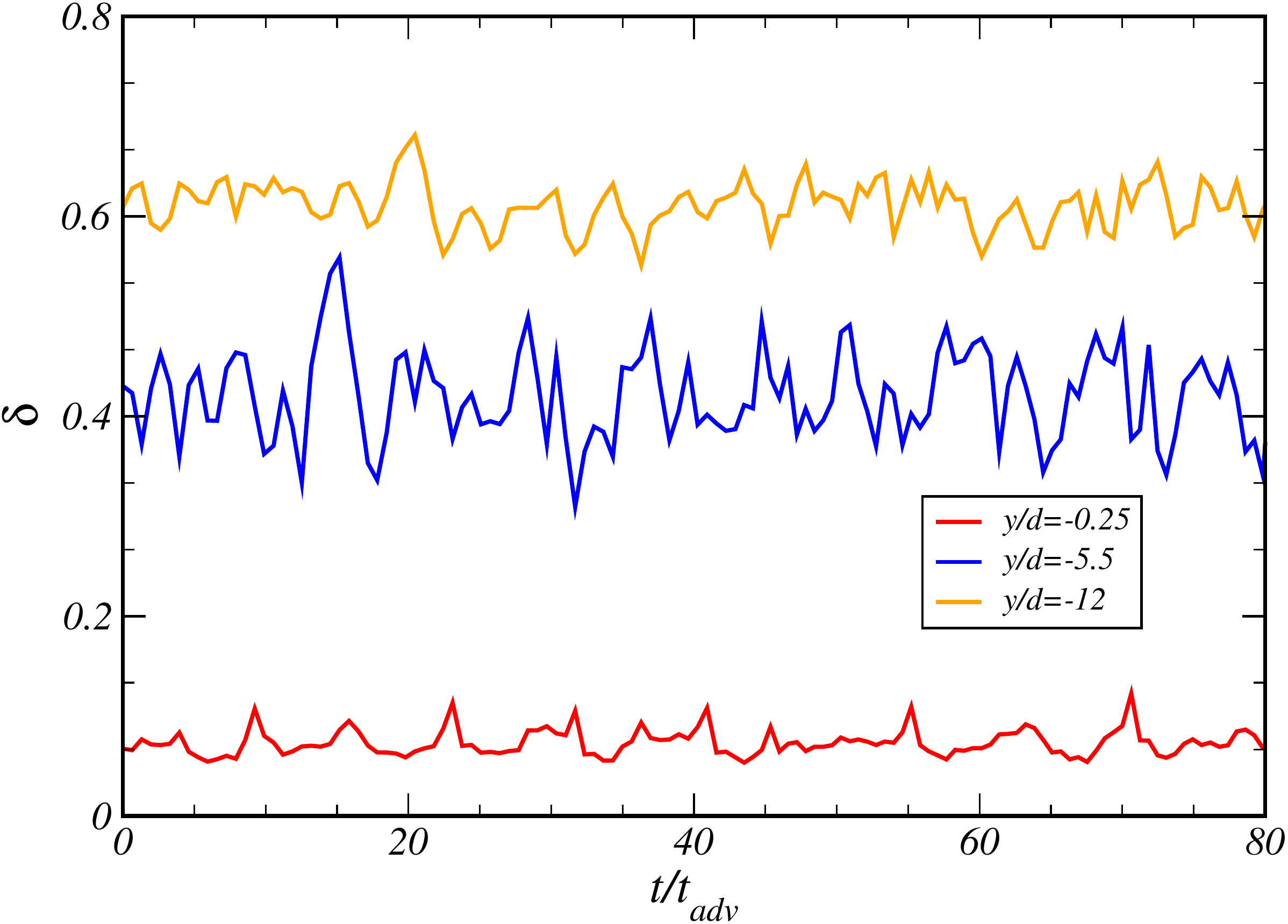} &
  	  \includegraphics[width=0.4\textwidth]{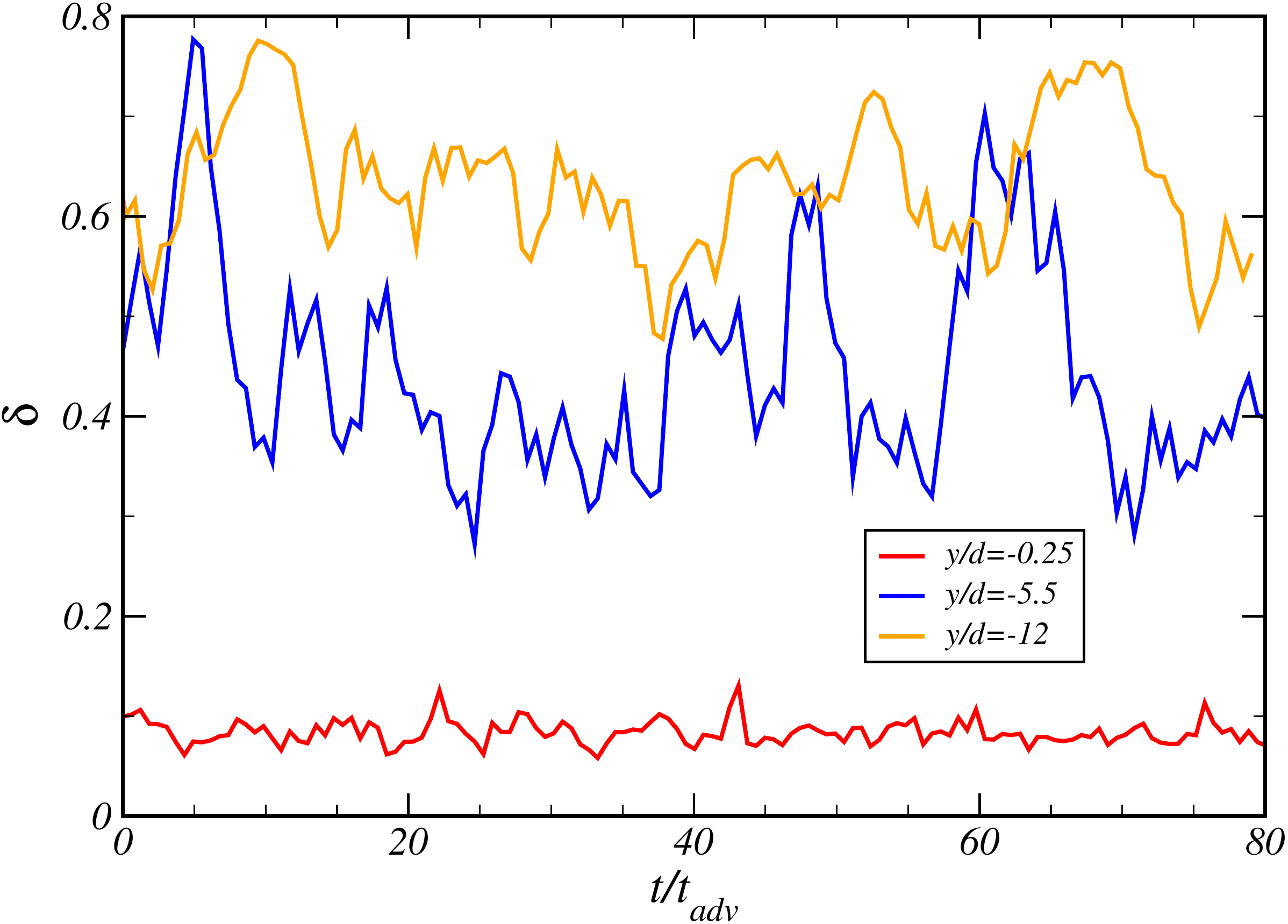}
	\end{tabular}
  \end{center}	
    \caption{Time evolution of mixing degree $\delta$ at $y/d=( -0.25, -5.5, -12)$ for WE (a) and WW (b) at $Re=1100$.}    
    \label{fig:Re1100DOM}
\end{figure}

\begin{figure}
  \begin{center}
    \begin{tabular}{ccc}
    (a) \small{$y/d=-0.25$} &
      (b) \small{$y/d=-5.5$} &(c) \small{$y/d=-12$} \\
    \includegraphics[width=0.25\textwidth]{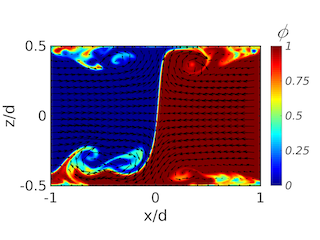} &
    \includegraphics[width=0.25\textwidth]{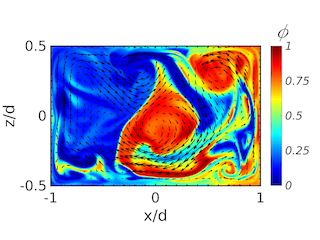} &
    \includegraphics[width=0.25\textwidth]{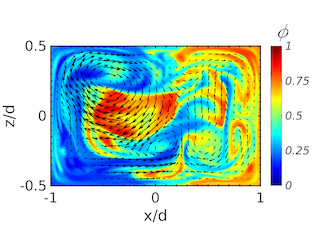} \\
    \includegraphics[width=0.25\textwidth]{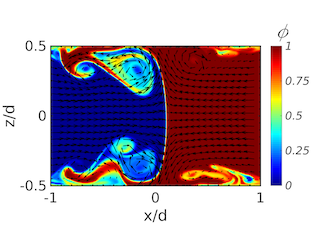} &
    \includegraphics[width=0.25\textwidth]{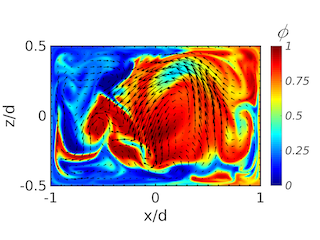} &
    \includegraphics[width=0.25\textwidth]{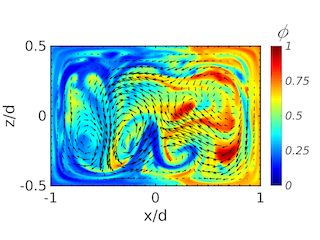} \\
    \includegraphics[width=0.25\textwidth]{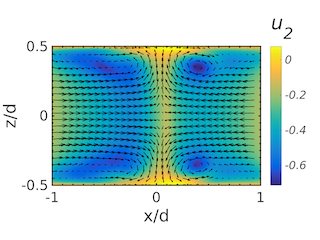} &
    \includegraphics[width=0.25\textwidth]{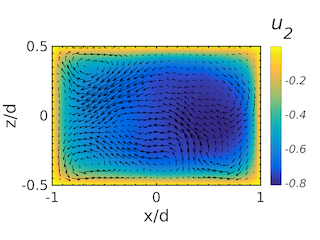} &
    \includegraphics[width=0.25\textwidth]{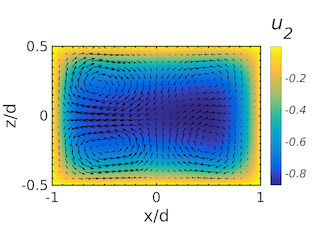} 
    \end{tabular}
  \end{center}
   \caption{Turbulent flow of WE at $Re=1100$. The top and middle row show snapshots of the concentration (colormap) and cross-sectional velocity field (arrows) at $t/t_{adv} \approx 0.0$ and $t/t_{adv} \approx 6.5$, respectively. The bottom row shows the time-averaged streamwise (colormap) and cross-sectional (arrows) velocities.}
   \label{fig:Re1100Cross}
\end{figure}

Fig.~\ref{fig:Re1100DOM} shows the time evolution of the cross-sectionally averaged mixing degree at $Re=1100$ in the turbulent regime. Here the flow at the inlets is still assumed laminar because flows in straight square ducts do not sustain a turbulent state for $Re<2200$ \cite{uhlmann2007marginally}. WW and WE have nearly the same mixing efficiency, but in WW the mixing degree fluctuates more strongly in time, as illustrated by the error bars of Fig.~\ref{fig:DarcyFrictionFactor}(a). The main flow features of WE are shown in Fig.~\ref{fig:Re1100Cross}. Near the junction the water and ethanol streams collide close to the center plane and the fluids are pressed towards the bottom and top wall as shown in the transversal velocities pointing in the $z$-direction. There, small irregular vortices arise and are transported along the mixing channel as they grow. Large fluctuations in the time evolution of $\delta$ show the raise of the ensuing large scale turbulent motions, whose  strong transversal velocities promote mixing as depicted in the top and middle panels of Fig.~\ref{fig:Re1100Cross}(b)--(c). For WE the local increase of viscosity damps these fluctuations earlier than for WW. Despite the spatio-temporal chaotic nature of the flow, some poorly mixed patches of fluid reach the outlet. 

At $Re=1100$ the flow is turbulent and hence the snapshots of the concentration in the top and middle rows of Fig.~\ref{fig:Re1100Cross} do not display symmetric flow structures. However, the top-down symmetry is preserved in average as shown in the bottom row of Fig.~\ref{fig:Re1100Cross}. Note that the average is not left-right symmetric because of the different shear stress at the two sides of the contact plane between ethanol and water. The average pattern exhibits a pair of two counter-rotating vortices, which are in fact quite similar to those of the steady vortex regime at low Reynolds number $Re<130$ \cite{Fani20136, thomas2010experimental, Orsi2013174}, see the top panels of Fig.~\ref{fig:WW160}. At $y/d=-5.5$, the mean vortical structures are streaked with small-scale motions, while, close to the outlet at $y/d=-12$, the flow has nearly recovered to a quasi-parabolic profile. The flow features of WW are very similar and hence not shown here. The main difference is that the time-averaged velocity profile is left-right symmetric because in contrast to WE, the viscosity has no dependence on the concentration.

%%%%%%%%%%%%%%%%%%%%%%%%%%%%%%%%%%%%%%%%%%%%%%
% Mixing 
%%%%%%%%%%%%%%%%%%%%%%%%%%%%%%%%%%%%%%%%%%%%%%
\section{Statistical analysis of mixing properties}

\begin{figure}
\begin{center}
\begin{tabular}{ccc}
(a) & (b) & \\
\includegraphics[width=0.5\textwidth]{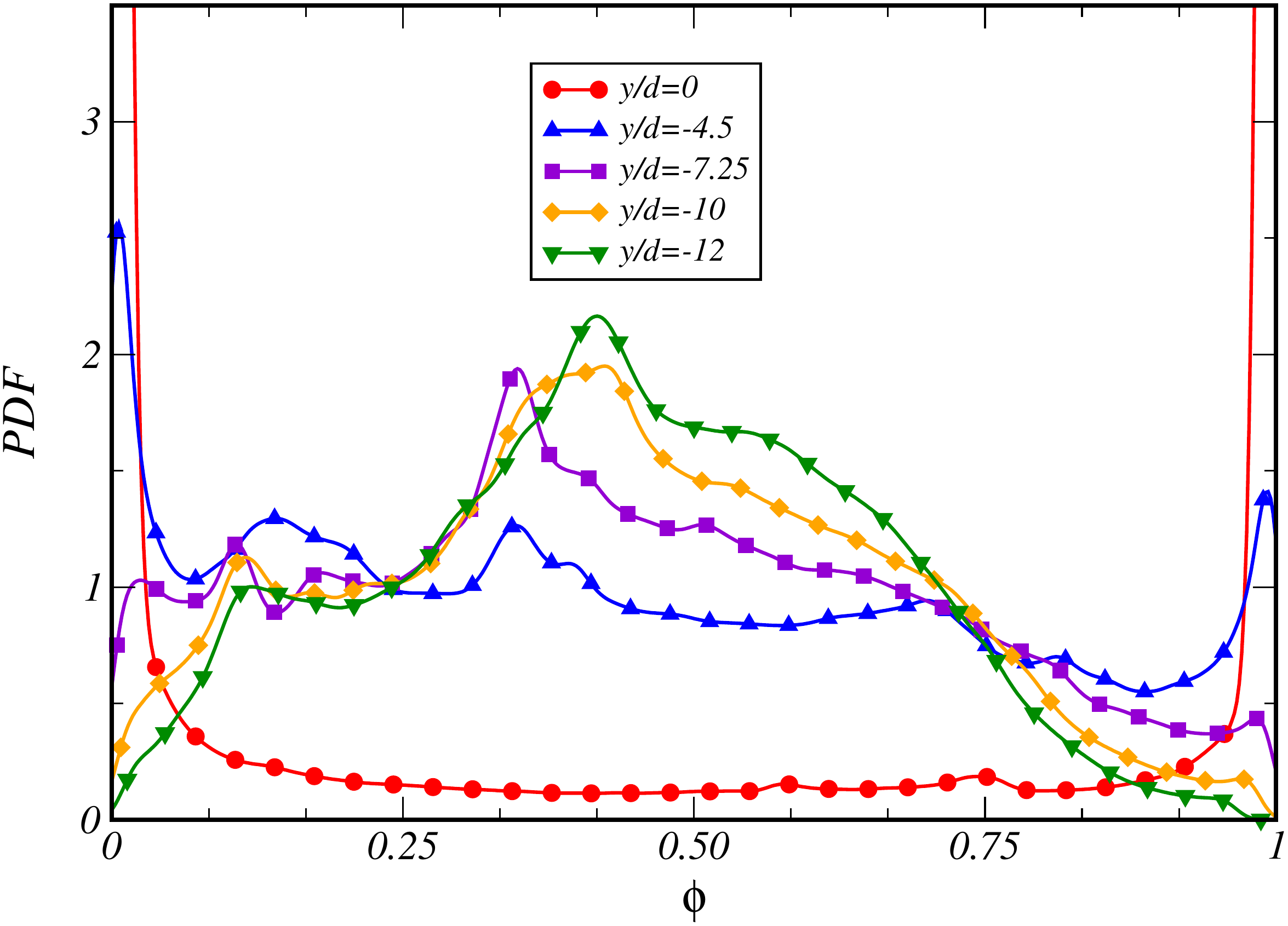}&
 \includegraphics[scale=3.2]{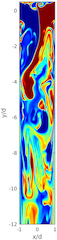}
\includegraphics[scale=3.2]{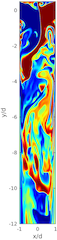}&
  \includegraphics[scale=0.2]{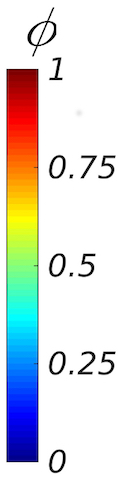}
\end{tabular}
\end{center}
 \caption{Stream-wise evolution of the mixing for $WE$ in the chaotic engulfment regime at $Re=425$. (a) Time-averaged probability density function of the concentration $\phi$ at selected cross sections in the mixing channel. (b) Snapshots of the concentration at mid-height ($z/d=0$) at time $t/t_{adv} \approx 0.0$ (left) and $t/t_{adv} \approx 3.2$ (right).}
   \label{fig:Re425PDFzplane}
\end{figure}
	% Peukert wants you to use different symbols for different curves, I totally agree

\begin{figure}
\begin{center}
\begin{tabular}{ccc}
(a) & (b) & \\
\includegraphics[width=0.5\textwidth]{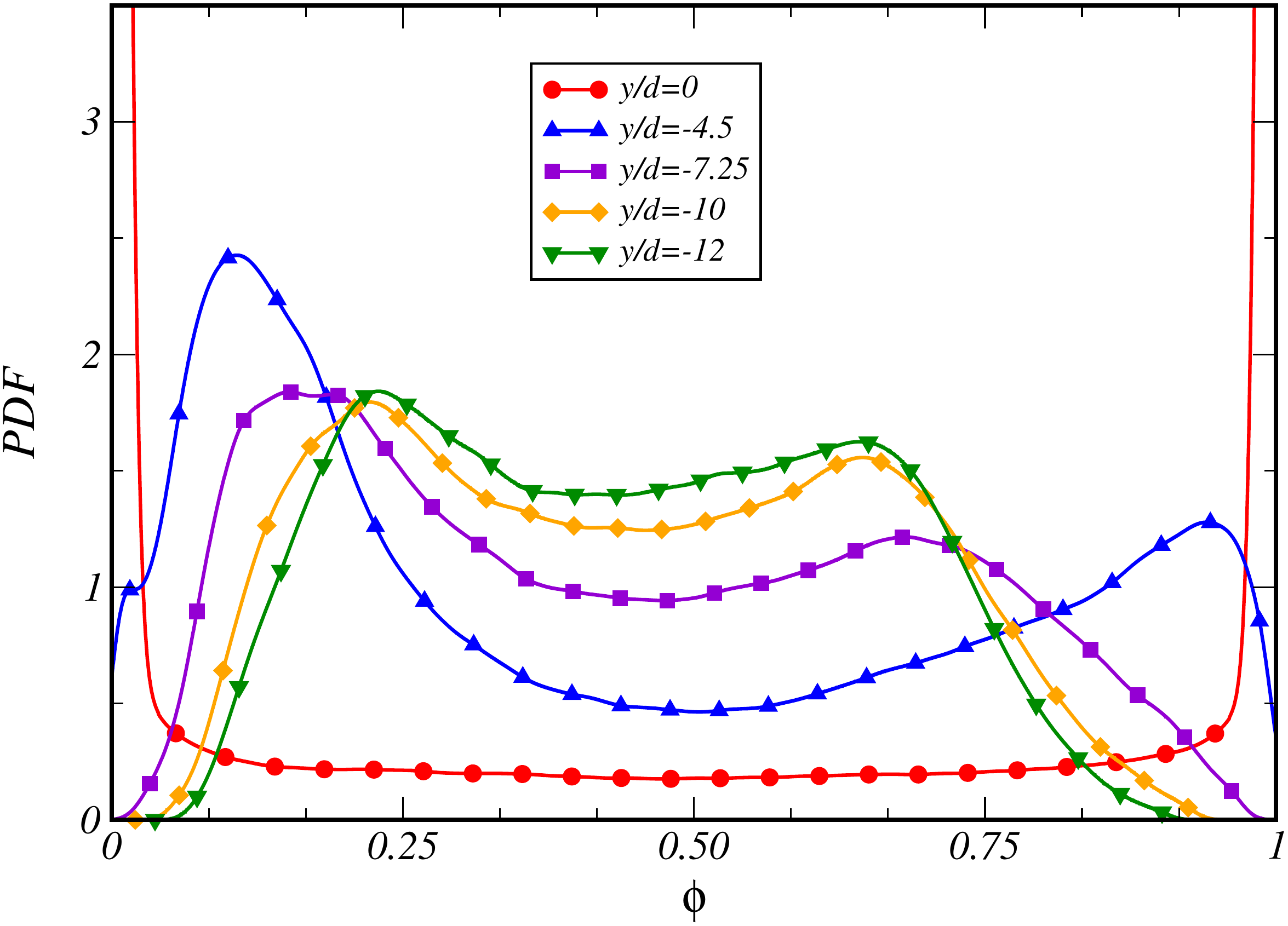}&
 \includegraphics[scale=3.2]{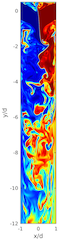}
  \includegraphics[scale=3.2]{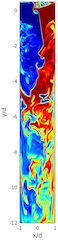}&
  \includegraphics[scale=0.2]{./colorbar.jpg}
\end{tabular}
\end{center}
 \caption{(a) Time-averaged probability density function of concentration $\phi$ for WE at $Re=1100$ and selected cross sections in the mixing channel. (b) Snapshots of colormap of concentration $\phi$ of WE for cross section $z/d=0$ at time $t/t_{adv} \approx 0.0 $ (left) and $t/t_{adv} \approx 6.5 $ (right) at $Re=1100$ corresponding to the time in Fig.~\ref{fig:Re1100Cross}.}
   \label{fig:Re1100PDFzplane}
\end{figure}
	% Peukert wants you to use different symbols for different curves, I totally agree

In the previous sections the mixing efficiency was evaluated with the mixing degree $\delta$ defined in equation \eqref{eqn:dom}, which entails a spatial-averaging over a channel cross-section. Hence $\delta$ cannot be used to infer the spatial homogeneity of the mixture, which is important for chemical reactions and precipitation processes \cite{Schwertfirm20071429,Gradl2006908,Wang2012252}. For the example of LAS precipitation, the driving force is supersaturation and this can vary by orders of magnitude depending on the local value of the concentration. In this section we briefly analyze the spatial homogeneity of the WE mixing process for the chaotic engulfment regime at $Re=425$ and the (symmetric) turbulent regime at $Re=1100$. Note that both cases exhibit a similar mixing degree at the outlet $\delta \approx 0.6$. 

Fig.~\ref{fig:Re425PDFzplane}(a) shows probability distribution functions (PDF) of the concentration at several cross-sections along the main channel for at $Re=425$. These PDFs were constructed by sampling a snapshot of a cross section every $0.25t/t_\text{adv}$ and then taking their time average. Fig.~\ref{fig:Re425PDFzplane}(b) shows two snapshots of the concentration at $z/d=0$ further illustrating the spatial evolution of the mixing process along the main channel. Between $y/d=-2$ and $-4$ the water and ethanol streams swap sides in the mixing channel, which is induced by the the growth of the Kelvin-Helmholtz vortices arising at the junction (see the top panel of Fig.~\ref{fig:WE425snaps}a). This is a salient feature of the engulfment regime and accounts for its higher mixing degree, when compared to the symmetric regime at the same Reynolds number (see the lower panel of Fig.~\ref{fig:WE425snaps}a). The intertwining of the fluid streams close to the junction result in a swift change from a bimodal to a plateau-like PDF. This can be interpreted following \citet{Schwertfirm20071429}, who found that the streamwise evolution of the PDF allows to determine the transport mechanisms dominating in the flow. In particular, a fast transition from a bimodal to plateau shape of the PDF, as seen in Fig.~\ref{fig:Re425PDFzplane}(a), implies that convective mixing is the dominant process, whereas an alteration in the plateau shape around $\phi \approx 0.5 $ points at mixing induced by molecular diffusion. Convective mixing is here ascribed to the stretching and folding of the contact plane between the fluids by vortices, which results in an increase of the flux of molecular diffusion. Toward the outlet the flow progressively relaminarises and molecular diffusion becomes the dominant process. In spite of the relatively high mixing degree at the outlet ($\delta=0.6$), there are many parcels of fluid leaving the channel practically unmixed, as illustrated by the dark blue and red regions close to the outlet in Fig.~\ref{fig:Re425PDFzplane}(b).Thus it appears that the vortex roll up in the engulfment regime results in rapid mixing but mainly around the vortex core. The vortices arising at the junction strongly rotate as the grow along the mixing channel but fail to involve the total mass of fluid because of their gradual viscous decay in the streamwise direction. 

It is useful to compare these results to the situation at $Re=1100$, for which the mean flow structures are top-down  symmetric and nearly left-right symmetric, as shown in Fig.~\ref{fig:Re1100Cross}. Despite the turbulent nature of the flow, Figure~\ref{fig:Re1100PDFzplane}(b) shows that the incoming water and ethanol streams flow nearly parallel to each other. Here convective mixing occurs solely because of fluctuating vortices acting across the interface, whereas the mean vortical pattern does barely contribute to mixing. This is in sharp contrast to the engulfment regime, for which most of the convective mixing is induced by the large-scale mean vortices swapping the streams immediately after the junction. This qualitative difference in the mixing process can be seen in the PDFs of Fig.~\ref{fig:Re1100PDFzplane}(a), which remain largely bimodal. The peaks on the water ($\phi=0$) and ethanol ($\phi=1$) sides shrink steadily while moving progressively toward a completely mixes state ($\phi = 0.5$). Beyond $ y/d=-10$ the fluctuating vortices lose gradually their intensity, albeit they are stronger than at $Re=425$ because of the lesser effect of viscosity, and there is a transition toward diffusion dominated mixing. 

%%%%%%%%%%%%%%%%%%%%%%%%%%%%%%%%%%%%%%%%%%%%%%
% Conclusion
%%%%%%%%%%%%%%%%%%%%%%%%%%%%%%%%%%%%%%%%%%%%%%
\section{Conclusion}
%%%%%%%%%%%%%%%%%%%%%%%%%%%%%%%%%%%%%%%%%%%%%%

Efficiently mixing fluids is key to many applications in chemical engineering. Hence it is desirable to develop simulation tools for detailed and accurate predictions of mixing processes, for example of their spatio-temporal homogeneity. In the case of two different miscible fluids, such as water and ethanol in LAS precipitation, the viscosity and density of the mixture depend strongly on the concentration and so the mixing process is particularly challenging to simulate. We designed and implemented an accurate numerical scheme to compute temporally and spatially resolved velocity and concentration fields in the turbulent regime. This is necessary for many applications, as in LAS precipitation, where particle formation depends on the local concentration distribution and its evolution along particle trajectories. Our scheme is based on a fully explicit low-storage Runge-Kutta method with decoupled transport equations and allows to simulate complex mixtures at the same computational cost as the transport of a passive scalar. Appropriate second-order spatial discretizations with flux limiters were adopted to handle large Schmidt numbers, here $Sc=600$, as usually occurring for miscible fluids. We demonstrated the potential of our methods by simulating water-water and water-ethanol mixtures for Reynolds numbers up to $Re=2000$ at a T-mixer. 

At low Reynolds numbers the flow is stratified and the mixing is purely diffusive. Here the two fluid streams flow parallel to each other, despite the presence of vortices confined to each of the streams. For the water-water case, this steady stratified vortex flow possess all the symmetries of the system, whereas in the water-ethanol case these symmetries are only approximate because of the complex dependence of the viscosity and density on the composition. As $Re$ increases the steady engulfment regime emerges. This is characterized by strong Kelvin-Helmholtz vortices at the junction, which break the left-right and up-down symmetries of the stratified flow. Nevertheless the flow remains symmetric with respect to a 180$^\circ$ rotation (again for WE the symmetry is only approximate). This qualitative change in the vortical pattern results in the convolution of the two streams along the main channel, which greatly enlarges the contact surface between them and thus enhances mixing considerably. As $Re$ further increases the flow becomes time-periodic and later chaotic and hence mixing continues to improve. Note that in average the rotationally symmetric large-scale vortices typical of the engulfment regime are preserved. Surprisingly, further increasing the Reynolds number results in a sudden reduction of the mixing efficiency, occurring for WW at $Re\approx 400$ and WE at $Re\approx 500$. This is caused by a transition to a distinct flow regime, which despite being time-periodic is top-down symmetric. In addition, near the junction it is nearly left-right symmetric and so the two fluid streams flow parallel to each other as in the stratified vortex flow at low Reynolds numbers. Here the mean vortical pattern contributes little to mixing and it is only through temporal fluctuations, i.e. vortices which act across the interface between the two fluids, that convective mixing occurs. As $Re$ increases, the fluctuations become larger but it is only at $Re=1100$ that the same mixing degree as for the chaotic engulfment at $Re=450$ is reached. However, it is worth noting that the mixing by temporal fluctuations at $Re=1100$ is spatially more homogenous than at $Re=450$, as in the latter many parcels of fluids leave the channel being practically unmixed. 

An interesting feature of this transition between the two flow types is the existence of a hysteresis region. Hence depending on the initial conditions or mode of operation, one state or the other can be realized, which actually results in a factor of three different mixing degree. Although a transition to a new more symmetric flow state had been previously identified by DNS \citep{Fani20147} as well as experiments  \citep{thomas2010experimental} in the water-water case, the finding of a hysteresis region and analysis of the mixing dynamics had not been reported. In agreement with previous simulations (\citep{Orsi2013174}), we found that the transition sequence occurs at larger $Re$ for WE than for WW, which implies that at the same $Re$ WW is better mixed. For the symmetric flow, however, this difference rapidly diminishes as $Re$ increases, which can be attributed to the increasing importance of turbulent fluctuations and also to water-ethanol mixture being only approximately symmetric in average. 

Our numerical approach, and potential improvements such as the multiple-resolution strategy of \cite{OstillaMonico2015308}, opens up the possibility to explore higher Reynolds number corresponding to real operating conditions with DNS, which will be pursued in the future. Our results contribute to understanding experimental measurements by providing detailed information of the velocity and concentration fields, and may be used as a reference to develop and test LES and RANS models. Finally, tracking the concentration and turbulent properties along flow trajectories will allow to deploy micromixing models and population balance equation solvers to predict the particle formation of LAS precipitation. This is part of our ongoing work. 

The authors would like to acknowledge the funding of the Deutsche Forschungsgemeinschaft (DFG) through the Cluster of Excellence Engineering of Advanced Materials (EAM) and Bayer Technology Services GmbH (BTS). Computing time from the Regionales Rechenzentrum Erlangen (RRZE) is acknowledged.

\bibliography{mybibfile}
\end{document}